%% file: TITELEI/dynamo-AMSA.tex
\input{./TITELEI/dynamic.top}
\documentclass{amsa} 

\usepackage{amsmath,     
            amsfonts,    
            amsthm,      
            textcomp,    
            graphics,    
            epsfig,      
            pstricks,      
            ifthen,
            verbatim,
            latexsym,
            exscale      
           }

\newif\ifamstheorem 
\amstheoremtrue
\newif\ifsptheorem 
\newif\ifnochaps
\nochapstrue
\usepackage{./LATEX/layout} 
\usepackage{./LATEX/math} 
\usepackage{pifont} 
\input{./LATEX/rgb.tex} 
\def\actcolor{black} 
\ifnocolor\renewcommand{\color}[1]{}\fi

\usepackage{./TITELEI/script}
\usepackage{./TITELEI/dynamic}

\begin{document}

\label{page:t}

\title{RELATIVISTIC ENTROPY INEQUALITY} 
\author{{H}ans {W}ilhelm {A}lt} 
\affiliation{Technische Universit\"at M\"unchen}
\email{alt@ma.tum.de}

\footcomment{
Physics and Astronomy Classification Scheme: 02.30.Jr, 04.20.Cv, 05.70.-a.
}

\maketitle

\input{./SOURCES/abstract-thermo-AMSA}

\newpage

\input{./SOURCES/content}
\input{SOURCES/refer-thermo}

\end{document}

%% file: LATEX/rgb.tex
  \definecolor{snow}{rgb}{1.000000,0.980392,0.980392}
  \definecolor{ghost white}{rgb}{0.972549,0.972549,1.000000}
  \definecolor{GhostWhite}{rgb}{0.972549,0.972549,1.000000}
  \definecolor{white smoke}{rgb}{0.960784,0.960784,0.960784}
  \definecolor{WhiteSmoke}{rgb}{0.960784,0.960784,0.960784}
  \definecolor{gainsboro}{rgb}{0.862745,0.862745,0.862745}
  \definecolor{floral white}{rgb}{1.000000,0.980392,0.941176}
  \definecolor{FloralWhite}{rgb}{1.000000,0.980392,0.941176}
  \definecolor{old lace}{rgb}{0.992157,0.960784,0.901961}
  \definecolor{OldLace}{rgb}{0.992157,0.960784,0.901961}
  \definecolor{linen}{rgb}{0.980392,0.941176,0.901961}
  \definecolor{antique white}{rgb}{0.980392,0.921569,0.843137}
  \definecolor{AntiqueWhite}{rgb}{0.980392,0.921569,0.843137}
  \definecolor{papaya whip}{rgb}{1.000000,0.937255,0.835294}
  \definecolor{PapayaWhip}{rgb}{1.000000,0.937255,0.835294}
  \definecolor{blanched almond}{rgb}{1.000000,0.921569,0.803922}
  \definecolor{BlanchedAlmond}{rgb}{1.000000,0.921569,0.803922}
  \definecolor{bisque}{rgb}{1.000000,0.894118,0.768627}
  \definecolor{peach puff}{rgb}{1.000000,0.854902,0.725490}
  \definecolor{PeachPuff}{rgb}{1.000000,0.854902,0.725490}
  \definecolor{navajo white}{rgb}{1.000000,0.870588,0.678431}
  \definecolor{NavajoWhite}{rgb}{1.000000,0.870588,0.678431}
  \definecolor{moccasin}{rgb}{1.000000,0.894118,0.709804}
  \definecolor{cornsilk}{rgb}{1.000000,0.972549,0.862745}
  \definecolor{ivory}{rgb}{1.000000,1.000000,0.941176}
  \definecolor{lemon chiffon}{rgb}{1.000000,0.980392,0.803922}
  \definecolor{LemonChiffon}{rgb}{1.000000,0.980392,0.803922}
  \definecolor{seashell}{rgb}{1.000000,0.960784,0.933333}
  \definecolor{honeydew}{rgb}{0.941176,1.000000,0.941176}
  \definecolor{mint cream}{rgb}{0.960784,1.000000,0.980392}
  \definecolor{MintCream}{rgb}{0.960784,1.000000,0.980392}
  \definecolor{azure}{rgb}{0.941176,1.000000,1.000000}
  \definecolor{alice blue}{rgb}{0.941176,0.972549,1.000000}
  \definecolor{AliceBlue}{rgb}{0.941176,0.972549,1.000000}
  \definecolor{lavender}{rgb}{0.901961,0.901961,0.980392}
  \definecolor{lavender blush}{rgb}{1.000000,0.941176,0.960784}
  \definecolor{LavenderBlush}{rgb}{1.000000,0.941176,0.960784}
  \definecolor{misty rose}{rgb}{1.000000,0.894118,0.882353}
  \definecolor{MistyRose}{rgb}{1.000000,0.894118,0.882353}
  \definecolor{white}{rgb}{1.000000,1.000000,1.000000}
  \definecolor{black}{rgb}{0.000000,0.000000,0.000000}
  \definecolor{dark slate gray}{rgb}{0.184314,0.309804,0.309804}
  \definecolor{DarkSlateGray}{rgb}{0.184314,0.309804,0.309804}
  \definecolor{dark slate grey}{rgb}{0.184314,0.309804,0.309804}
  \definecolor{DarkSlateGrey}{rgb}{0.184314,0.309804,0.309804}
  \definecolor{dim gray}{rgb}{0.411765,0.411765,0.411765}
  \definecolor{DimGray}{rgb}{0.411765,0.411765,0.411765}
  \definecolor{dim grey}{rgb}{0.411765,0.411765,0.411765}
  \definecolor{DimGrey}{rgb}{0.411765,0.411765,0.411765}
  \definecolor{slate gray}{rgb}{0.439216,0.501961,0.564706}
  \definecolor{SlateGray}{rgb}{0.439216,0.501961,0.564706}
  \definecolor{slate grey}{rgb}{0.439216,0.501961,0.564706}
  \definecolor{SlateGrey}{rgb}{0.439216,0.501961,0.564706}
  \definecolor{light slate gray}{rgb}{0.466667,0.533333,0.600000}
  \definecolor{LightSlateGray}{rgb}{0.466667,0.533333,0.600000}
  \definecolor{light slate grey}{rgb}{0.466667,0.533333,0.600000}
  \definecolor{LightSlateGrey}{rgb}{0.466667,0.533333,0.600000}
  \definecolor{gray}{rgb}{0.745098,0.745098,0.745098}
  \definecolor{grey}{rgb}{0.745098,0.745098,0.745098}
  \definecolor{light grey}{rgb}{0.827451,0.827451,0.827451}
  \definecolor{LightGrey}{rgb}{0.827451,0.827451,0.827451}
  \definecolor{light gray}{rgb}{0.827451,0.827451,0.827451}
  \definecolor{LightGray}{rgb}{0.827451,0.827451,0.827451}
  \definecolor{midnight blue}{rgb}{0.098039,0.098039,0.439216}
  \definecolor{MidnightBlue}{rgb}{0.098039,0.098039,0.439216}
  \definecolor{navy}{rgb}{0.000000,0.000000,0.501961}
  \definecolor{navy blue}{rgb}{0.000000,0.000000,0.501961}
  \definecolor{NavyBlue}{rgb}{0.000000,0.000000,0.501961}
  \definecolor{cornflower blue}{rgb}{0.392157,0.584314,0.929412}
  \definecolor{CornflowerBlue}{rgb}{0.392157,0.584314,0.929412}
  \definecolor{dark slate blue}{rgb}{0.282353,0.239216,0.545098}
  \definecolor{DarkSlateBlue}{rgb}{0.282353,0.239216,0.545098}
  \definecolor{slate blue}{rgb}{0.415686,0.352941,0.803922}
  \definecolor{SlateBlue}{rgb}{0.415686,0.352941,0.803922}
  \definecolor{medium slate blue}{rgb}{0.482353,0.407843,0.933333}
  \definecolor{MediumSlateBlue}{rgb}{0.482353,0.407843,0.933333}
  \definecolor{light slate blue}{rgb}{0.517647,0.439216,1.000000}
  \definecolor{LightSlateBlue}{rgb}{0.517647,0.439216,1.000000}
  \definecolor{medium blue}{rgb}{0.000000,0.000000,0.803922}
  \definecolor{MediumBlue}{rgb}{0.000000,0.000000,0.803922}
  \definecolor{royal blue}{rgb}{0.254902,0.411765,0.882353}
  \definecolor{RoyalBlue}{rgb}{0.254902,0.411765,0.882353}
  \definecolor{blue}{rgb}{0.000000,0.000000,1.000000}
  \definecolor{dodger blue}{rgb}{0.117647,0.564706,1.000000}
  \definecolor{DodgerBlue}{rgb}{0.117647,0.564706,1.000000}
  \definecolor{deep sky blue}{rgb}{0.000000,0.749020,1.000000}
  \definecolor{DeepSkyBlue}{rgb}{0.000000,0.749020,1.000000}
  \definecolor{sky blue}{rgb}{0.529412,0.807843,0.921569}
  \definecolor{SkyBlue}{rgb}{0.529412,0.807843,0.921569}
  \definecolor{light sky blue}{rgb}{0.529412,0.807843,0.980392}
  \definecolor{LightSkyBlue}{rgb}{0.529412,0.807843,0.980392}
  \definecolor{steel blue}{rgb}{0.274510,0.509804,0.705882}
  \definecolor{SteelBlue}{rgb}{0.274510,0.509804,0.705882}
  \definecolor{light steel blue}{rgb}{0.690196,0.768627,0.870588}
  \definecolor{LightSteelBlue}{rgb}{0.690196,0.768627,0.870588}
  \definecolor{light blue}{rgb}{0.678431,0.847059,0.901961}
  \definecolor{LightBlue}{rgb}{0.678431,0.847059,0.901961}
  \definecolor{powder blue}{rgb}{0.690196,0.878431,0.901961}
  \definecolor{PowderBlue}{rgb}{0.690196,0.878431,0.901961}
  \definecolor{pale turquoise}{rgb}{0.686275,0.933333,0.933333}
  \definecolor{PaleTurquoise}{rgb}{0.686275,0.933333,0.933333}
  \definecolor{dark turquoise}{rgb}{0.000000,0.807843,0.819608}
  \definecolor{DarkTurquoise}{rgb}{0.000000,0.807843,0.819608}
  \definecolor{medium turquoise}{rgb}{0.282353,0.819608,0.800000}
  \definecolor{MediumTurquoise}{rgb}{0.282353,0.819608,0.800000}
  \definecolor{turquoise}{rgb}{0.250980,0.878431,0.815686}
  \definecolor{cyan}{rgb}{0.000000,1.000000,1.000000}
  \definecolor{light cyan}{rgb}{0.878431,1.000000,1.000000}
  \definecolor{LightCyan}{rgb}{0.878431,1.000000,1.000000}
  \definecolor{cadet blue}{rgb}{0.372549,0.619608,0.627451}
  \definecolor{CadetBlue}{rgb}{0.372549,0.619608,0.627451}
  \definecolor{medium aquamarine}{rgb}{0.400000,0.803922,0.666667}
  \definecolor{MediumAquamarine}{rgb}{0.400000,0.803922,0.666667}
  \definecolor{aquamarine}{rgb}{0.498039,1.000000,0.831373}
  \definecolor{dark green}{rgb}{0.000000,0.392157,0.000000}
  \definecolor{DarkGreen}{rgb}{0.000000,0.392157,0.000000}
  \definecolor{dark olive green}{rgb}{0.333333,0.419608,0.184314}
  \definecolor{DarkOliveGreen}{rgb}{0.333333,0.419608,0.184314}
  \definecolor{dark sea green}{rgb}{0.560784,0.737255,0.560784}
  \definecolor{DarkSeaGreen}{rgb}{0.560784,0.737255,0.560784}
  \definecolor{sea green}{rgb}{0.180392,0.545098,0.341176}
  \definecolor{SeaGreen}{rgb}{0.180392,0.545098,0.341176}
  \definecolor{medium sea green}{rgb}{0.235294,0.701961,0.443137}
  \definecolor{MediumSeaGreen}{rgb}{0.235294,0.701961,0.443137}
  \definecolor{light sea green}{rgb}{0.125490,0.698039,0.666667}
  \definecolor{LightSeaGreen}{rgb}{0.125490,0.698039,0.666667}
  \definecolor{pale green}{rgb}{0.596078,0.984314,0.596078}
  \definecolor{PaleGreen}{rgb}{0.596078,0.984314,0.596078}
  \definecolor{spring green}{rgb}{0.000000,1.000000,0.498039}
  \definecolor{SpringGreen}{rgb}{0.000000,1.000000,0.498039}
  \definecolor{lawn green}{rgb}{0.486275,0.988235,0.000000}
  \definecolor{LawnGreen}{rgb}{0.486275,0.988235,0.000000}
  \definecolor{green}{rgb}{0.000000,1.000000,0.000000}
  \definecolor{chartreuse}{rgb}{0.498039,1.000000,0.000000}
  \definecolor{medium spring green}{rgb}{0.000000,0.980392,0.603922}
  \definecolor{MediumSpringGreen}{rgb}{0.000000,0.980392,0.603922}
  \definecolor{green yellow}{rgb}{0.678431,1.000000,0.184314}
  \definecolor{GreenYellow}{rgb}{0.678431,1.000000,0.184314}
  \definecolor{lime green}{rgb}{0.196078,0.803922,0.196078}
  \definecolor{LimeGreen}{rgb}{0.196078,0.803922,0.196078}
  \definecolor{yellow green}{rgb}{0.603922,0.803922,0.196078}
  \definecolor{YellowGreen}{rgb}{0.603922,0.803922,0.196078}
  \definecolor{forest green}{rgb}{0.133333,0.545098,0.133333}
  \definecolor{ForestGreen}{rgb}{0.133333,0.545098,0.133333}
  \definecolor{olive drab}{rgb}{0.419608,0.556863,0.137255}
  \definecolor{OliveDrab}{rgb}{0.419608,0.556863,0.137255}
  \definecolor{dark khaki}{rgb}{0.741176,0.717647,0.419608}
  \definecolor{DarkKhaki}{rgb}{0.741176,0.717647,0.419608}
  \definecolor{khaki}{rgb}{0.941176,0.901961,0.549020}
  \definecolor{pale goldenrod}{rgb}{0.933333,0.909804,0.666667}
  \definecolor{PaleGoldenrod}{rgb}{0.933333,0.909804,0.666667}
  \definecolor{light goldenrod yellow}{rgb}{0.980392,0.980392,0.823529}
  \definecolor{LightGoldenrodYellow}{rgb}{0.980392,0.980392,0.823529}
  \definecolor{light yellow}{rgb}{1.000000,1.000000,0.878431}
  \definecolor{LightYellow}{rgb}{1.000000,1.000000,0.878431}
  \definecolor{yellow}{rgb}{1.000000,1.000000,0.000000}
  \definecolor{gold}{rgb}{1.000000,0.843137,0.000000}
  \definecolor{light goldenrod}{rgb}{0.933333,0.866667,0.509804}
  \definecolor{LightGoldenrod}{rgb}{0.933333,0.866667,0.509804}
  \definecolor{goldenrod}{rgb}{0.854902,0.647059,0.125490}
  \definecolor{dark goldenrod}{rgb}{0.721569,0.525490,0.043137}
  \definecolor{DarkGoldenrod}{rgb}{0.721569,0.525490,0.043137}
  \definecolor{rosy brown}{rgb}{0.737255,0.560784,0.560784}
  \definecolor{RosyBrown}{rgb}{0.737255,0.560784,0.560784}
  \definecolor{indian red}{rgb}{0.803922,0.360784,0.360784}
  \definecolor{IndianRed}{rgb}{0.803922,0.360784,0.360784}
  \definecolor{saddle brown}{rgb}{0.545098,0.270588,0.074510}
  \definecolor{SaddleBrown}{rgb}{0.545098,0.270588,0.074510}
  \definecolor{sienna}{rgb}{0.627451,0.321569,0.176471}
  \definecolor{peru}{rgb}{0.803922,0.521569,0.247059}
  \definecolor{burlywood}{rgb}{0.870588,0.721569,0.529412}
  \definecolor{beige}{rgb}{0.960784,0.960784,0.862745}
  \definecolor{wheat}{rgb}{0.960784,0.870588,0.701961}
  \definecolor{sandy brown}{rgb}{0.956863,0.643137,0.376471}
  \definecolor{SandyBrown}{rgb}{0.956863,0.643137,0.376471}
  \definecolor{tan}{rgb}{0.823529,0.705882,0.549020}
  \definecolor{chocolate}{rgb}{0.823529,0.411765,0.117647}
  \definecolor{firebrick}{rgb}{0.698039,0.133333,0.133333}
  \definecolor{brown}{rgb}{0.647059,0.164706,0.164706}
  \definecolor{dark salmon}{rgb}{0.913725,0.588235,0.478431}
  \definecolor{DarkSalmon}{rgb}{0.913725,0.588235,0.478431}
  \definecolor{salmon}{rgb}{0.980392,0.501961,0.447059}
  \definecolor{light salmon}{rgb}{1.000000,0.627451,0.478431}
  \definecolor{LightSalmon}{rgb}{1.000000,0.627451,0.478431}
  \definecolor{orange}{rgb}{1.000000,0.647059,0.000000}
  \definecolor{dark orange}{rgb}{1.000000,0.549020,0.000000}
  \definecolor{DarkOrange}{rgb}{1.000000,0.549020,0.000000}
  \definecolor{coral}{rgb}{1.000000,0.498039,0.313726}
  \definecolor{light coral}{rgb}{0.941176,0.501961,0.501961}
  \definecolor{LightCoral}{rgb}{0.941176,0.501961,0.501961}
  \definecolor{tomato}{rgb}{1.000000,0.388235,0.278431}
  \definecolor{orange red}{rgb}{1.000000,0.270588,0.000000}
  \definecolor{OrangeRed}{rgb}{1.000000,0.270588,0.000000}
  \definecolor{red}{rgb}{1.000000,0.000000,0.000000}
  \definecolor{hot pink}{rgb}{1.000000,0.411765,0.705882}
  \definecolor{HotPink}{rgb}{1.000000,0.411765,0.705882}
  \definecolor{deep pink}{rgb}{1.000000,0.078431,0.576471}
  \definecolor{DeepPink}{rgb}{1.000000,0.078431,0.576471}
  \definecolor{pink}{rgb}{1.000000,0.752941,0.796078}
  \definecolor{light pink}{rgb}{1.000000,0.713726,0.756863}
  \definecolor{LightPink}{rgb}{1.000000,0.713726,0.756863}
  \definecolor{pale violet red}{rgb}{0.858824,0.439216,0.576471}
  \definecolor{PaleVioletRed}{rgb}{0.858824,0.439216,0.576471}
  \definecolor{maroon}{rgb}{0.690196,0.188235,0.376471}
  \definecolor{medium violet red}{rgb}{0.780392,0.082353,0.521569}
  \definecolor{MediumVioletRed}{rgb}{0.780392,0.082353,0.521569}
  \definecolor{violet red}{rgb}{0.815686,0.125490,0.564706}
  \definecolor{VioletRed}{rgb}{0.815686,0.125490,0.564706}
  \definecolor{magenta}{rgb}{1.000000,0.000000,1.000000}
  \definecolor{violet}{rgb}{0.933333,0.509804,0.933333}
  \definecolor{plum}{rgb}{0.866667,0.627451,0.866667}
  \definecolor{orchid}{rgb}{0.854902,0.439216,0.839216}
  \definecolor{medium orchid}{rgb}{0.729412,0.333333,0.827451}
  \definecolor{MediumOrchid}{rgb}{0.729412,0.333333,0.827451}
  \definecolor{dark orchid}{rgb}{0.600000,0.196078,0.800000}
  \definecolor{DarkOrchid}{rgb}{0.600000,0.196078,0.800000}
  \definecolor{dark violet}{rgb}{0.580392,0.000000,0.827451}
  \definecolor{DarkViolet}{rgb}{0.580392,0.000000,0.827451}
  \definecolor{blue violet}{rgb}{0.541176,0.168627,0.886275}
  \definecolor{BlueViolet}{rgb}{0.541176,0.168627,0.886275}
  \definecolor{purple}{rgb}{0.627451,0.125490,0.941176}
  \definecolor{medium purple}{rgb}{0.576471,0.439216,0.858824}
  \definecolor{MediumPurple}{rgb}{0.576471,0.439216,0.858824}
  \definecolor{thistle}{rgb}{0.847059,0.749020,0.847059}
  \definecolor{snow1}{rgb}{1.000000,0.980392,0.980392}
  \definecolor{snow2}{rgb}{0.933333,0.913725,0.913725}
  \definecolor{snow3}{rgb}{0.803922,0.788235,0.788235}
  \definecolor{snow4}{rgb}{0.545098,0.537255,0.537255}
  \definecolor{seashell1}{rgb}{1.000000,0.960784,0.933333}
  \definecolor{seashell2}{rgb}{0.933333,0.898039,0.870588}
  \definecolor{seashell3}{rgb}{0.803922,0.772549,0.749020}
  \definecolor{seashell4}{rgb}{0.545098,0.525490,0.509804}
  \definecolor{AntiqueWhite1}{rgb}{1.000000,0.937255,0.858824}
  \definecolor{AntiqueWhite2}{rgb}{0.933333,0.874510,0.800000}
  \definecolor{AntiqueWhite3}{rgb}{0.803922,0.752941,0.690196}
  \definecolor{AntiqueWhite4}{rgb}{0.545098,0.513726,0.470588}
  \definecolor{bisque1}{rgb}{1.000000,0.894118,0.768627}
  \definecolor{bisque2}{rgb}{0.933333,0.835294,0.717647}
  \definecolor{bisque3}{rgb}{0.803922,0.717647,0.619608}
  \definecolor{bisque4}{rgb}{0.545098,0.490196,0.419608}
  \definecolor{PeachPuff1}{rgb}{1.000000,0.854902,0.725490}
  \definecolor{PeachPuff2}{rgb}{0.933333,0.796078,0.678431}
  \definecolor{PeachPuff3}{rgb}{0.803922,0.686275,0.584314}
  \definecolor{PeachPuff4}{rgb}{0.545098,0.466667,0.396078}
  \definecolor{NavajoWhite1}{rgb}{1.000000,0.870588,0.678431}
  \definecolor{NavajoWhite2}{rgb}{0.933333,0.811765,0.631373}
  \definecolor{NavajoWhite3}{rgb}{0.803922,0.701961,0.545098}
  \definecolor{NavajoWhite4}{rgb}{0.545098,0.474510,0.368627}
  \definecolor{LemonChiffon1}{rgb}{1.000000,0.980392,0.803922}
  \definecolor{LemonChiffon2}{rgb}{0.933333,0.913725,0.749020}
  \definecolor{LemonChiffon3}{rgb}{0.803922,0.788235,0.647059}
  \definecolor{LemonChiffon4}{rgb}{0.545098,0.537255,0.439216}
  \definecolor{cornsilk1}{rgb}{1.000000,0.972549,0.862745}
  \definecolor{cornsilk2}{rgb}{0.933333,0.909804,0.803922}
  \definecolor{cornsilk3}{rgb}{0.803922,0.784314,0.694118}
  \definecolor{cornsilk4}{rgb}{0.545098,0.533333,0.470588}
  \definecolor{ivory1}{rgb}{1.000000,1.000000,0.941176}
  \definecolor{ivory2}{rgb}{0.933333,0.933333,0.878431}
  \definecolor{ivory3}{rgb}{0.803922,0.803922,0.756863}
  \definecolor{ivory4}{rgb}{0.545098,0.545098,0.513726}
  \definecolor{honeydew1}{rgb}{0.941176,1.000000,0.941176}
  \definecolor{honeydew2}{rgb}{0.878431,0.933333,0.878431}
  \definecolor{honeydew3}{rgb}{0.756863,0.803922,0.756863}
  \definecolor{honeydew4}{rgb}{0.513726,0.545098,0.513726}
  \definecolor{LavenderBlush1}{rgb}{1.000000,0.941176,0.960784}
  \definecolor{LavenderBlush2}{rgb}{0.933333,0.878431,0.898039}
  \definecolor{LavenderBlush3}{rgb}{0.803922,0.756863,0.772549}
  \definecolor{LavenderBlush4}{rgb}{0.545098,0.513726,0.525490}
  \definecolor{MistyRose1}{rgb}{1.000000,0.894118,0.882353}
  \definecolor{MistyRose2}{rgb}{0.933333,0.835294,0.823529}
  \definecolor{MistyRose3}{rgb}{0.803922,0.717647,0.709804}
  \definecolor{MistyRose4}{rgb}{0.545098,0.490196,0.482353}
  \definecolor{azure1}{rgb}{0.941176,1.000000,1.000000}
  \definecolor{azure2}{rgb}{0.878431,0.933333,0.933333}
  \definecolor{azure3}{rgb}{0.756863,0.803922,0.803922}
  \definecolor{azure4}{rgb}{0.513726,0.545098,0.545098}
  \definecolor{SlateBlue1}{rgb}{0.513726,0.435294,1.000000}
  \definecolor{SlateBlue2}{rgb}{0.478431,0.403922,0.933333}
  \definecolor{SlateBlue3}{rgb}{0.411765,0.349020,0.803922}
  \definecolor{SlateBlue4}{rgb}{0.278431,0.235294,0.545098}
  \definecolor{RoyalBlue1}{rgb}{0.282353,0.462745,1.000000}
  \definecolor{RoyalBlue2}{rgb}{0.262745,0.431373,0.933333}
  \definecolor{RoyalBlue3}{rgb}{0.227451,0.372549,0.803922}
  \definecolor{RoyalBlue4}{rgb}{0.152941,0.250980,0.545098}
  \definecolor{blue1}{rgb}{0.000000,0.000000,1.000000}
  \definecolor{blue2}{rgb}{0.000000,0.000000,0.933333}
  \definecolor{blue3}{rgb}{0.000000,0.000000,0.803922}
  \definecolor{blue4}{rgb}{0.000000,0.000000,0.545098}
  \definecolor{DodgerBlue1}{rgb}{0.117647,0.564706,1.000000}
  \definecolor{DodgerBlue2}{rgb}{0.109804,0.525490,0.933333}
  \definecolor{DodgerBlue3}{rgb}{0.094118,0.454902,0.803922}
  \definecolor{DodgerBlue4}{rgb}{0.062745,0.305882,0.545098}
  \definecolor{SteelBlue1}{rgb}{0.388235,0.721569,1.000000}
  \definecolor{SteelBlue2}{rgb}{0.360784,0.674510,0.933333}
  \definecolor{SteelBlue3}{rgb}{0.309804,0.580392,0.803922}
  \definecolor{SteelBlue4}{rgb}{0.211765,0.392157,0.545098}
  \definecolor{DeepSkyBlue1}{rgb}{0.000000,0.749020,1.000000}
  \definecolor{DeepSkyBlue2}{rgb}{0.000000,0.698039,0.933333}
  \definecolor{DeepSkyBlue3}{rgb}{0.000000,0.603922,0.803922}
  \definecolor{DeepSkyBlue4}{rgb}{0.000000,0.407843,0.545098}
  \definecolor{SkyBlue1}{rgb}{0.529412,0.807843,1.000000}
  \definecolor{SkyBlue2}{rgb}{0.494118,0.752941,0.933333}
  \definecolor{SkyBlue3}{rgb}{0.423529,0.650980,0.803922}
  \definecolor{SkyBlue4}{rgb}{0.290196,0.439216,0.545098}
  \definecolor{LightSkyBlue1}{rgb}{0.690196,0.886275,1.000000}
  \definecolor{LightSkyBlue2}{rgb}{0.643137,0.827451,0.933333}
  \definecolor{LightSkyBlue3}{rgb}{0.552941,0.713726,0.803922}
  \definecolor{LightSkyBlue4}{rgb}{0.376471,0.482353,0.545098}
  \definecolor{SlateGray1}{rgb}{0.776471,0.886275,1.000000}
  \definecolor{SlateGray2}{rgb}{0.725490,0.827451,0.933333}
  \definecolor{SlateGray3}{rgb}{0.623529,0.713726,0.803922}
  \definecolor{SlateGray4}{rgb}{0.423529,0.482353,0.545098}
  \definecolor{LightSteelBlue1}{rgb}{0.792157,0.882353,1.000000}
  \definecolor{LightSteelBlue2}{rgb}{0.737255,0.823529,0.933333}
  \definecolor{LightSteelBlue3}{rgb}{0.635294,0.709804,0.803922}
  \definecolor{LightSteelBlue4}{rgb}{0.431373,0.482353,0.545098}
  \definecolor{LightBlue1}{rgb}{0.749020,0.937255,1.000000}
  \definecolor{LightBlue2}{rgb}{0.698039,0.874510,0.933333}
  \definecolor{LightBlue3}{rgb}{0.603922,0.752941,0.803922}
  \definecolor{LightBlue4}{rgb}{0.407843,0.513726,0.545098}
  \definecolor{LightCyan1}{rgb}{0.878431,1.000000,1.000000}
  \definecolor{LightCyan2}{rgb}{0.819608,0.933333,0.933333}
  \definecolor{LightCyan3}{rgb}{0.705882,0.803922,0.803922}
  \definecolor{LightCyan4}{rgb}{0.478431,0.545098,0.545098}
  \definecolor{PaleTurquoise1}{rgb}{0.733333,1.000000,1.000000}
  \definecolor{PaleTurquoise2}{rgb}{0.682353,0.933333,0.933333}
  \definecolor{PaleTurquoise3}{rgb}{0.588235,0.803922,0.803922}
  \definecolor{PaleTurquoise4}{rgb}{0.400000,0.545098,0.545098}
  \definecolor{CadetBlue1}{rgb}{0.596078,0.960784,1.000000}
  \definecolor{CadetBlue2}{rgb}{0.556863,0.898039,0.933333}
  \definecolor{CadetBlue3}{rgb}{0.478431,0.772549,0.803922}
  \definecolor{CadetBlue4}{rgb}{0.325490,0.525490,0.545098}
  \definecolor{turquoise1}{rgb}{0.000000,0.960784,1.000000}
  \definecolor{turquoise2}{rgb}{0.000000,0.898039,0.933333}
  \definecolor{turquoise3}{rgb}{0.000000,0.772549,0.803922}
  \definecolor{turquoise4}{rgb}{0.000000,0.525490,0.545098}
  \definecolor{cyan1}{rgb}{0.000000,1.000000,1.000000}
  \definecolor{cyan2}{rgb}{0.000000,0.933333,0.933333}
  \definecolor{cyan3}{rgb}{0.000000,0.803922,0.803922}
  \definecolor{cyan4}{rgb}{0.000000,0.545098,0.545098}
  \definecolor{DarkSlateGray1}{rgb}{0.592157,1.000000,1.000000}
  \definecolor{DarkSlateGray2}{rgb}{0.552941,0.933333,0.933333}
  \definecolor{DarkSlateGray3}{rgb}{0.474510,0.803922,0.803922}
  \definecolor{DarkSlateGray4}{rgb}{0.321569,0.545098,0.545098}
  \definecolor{aquamarine1}{rgb}{0.498039,1.000000,0.831373}
  \definecolor{aquamarine2}{rgb}{0.462745,0.933333,0.776471}
  \definecolor{aquamarine3}{rgb}{0.400000,0.803922,0.666667}
  \definecolor{aquamarine4}{rgb}{0.270588,0.545098,0.454902}
  \definecolor{DarkSeaGreen1}{rgb}{0.756863,1.000000,0.756863}
  \definecolor{DarkSeaGreen2}{rgb}{0.705882,0.933333,0.705882}
  \definecolor{DarkSeaGreen3}{rgb}{0.607843,0.803922,0.607843}
  \definecolor{DarkSeaGreen4}{rgb}{0.411765,0.545098,0.411765}
  \definecolor{SeaGreen1}{rgb}{0.329412,1.000000,0.623529}
  \definecolor{SeaGreen2}{rgb}{0.305882,0.933333,0.580392}
  \definecolor{SeaGreen3}{rgb}{0.262745,0.803922,0.501961}
  \definecolor{SeaGreen4}{rgb}{0.180392,0.545098,0.341176}
  \definecolor{PaleGreen1}{rgb}{0.603922,1.000000,0.603922}
  \definecolor{PaleGreen2}{rgb}{0.564706,0.933333,0.564706}
  \definecolor{PaleGreen3}{rgb}{0.486275,0.803922,0.486275}
  \definecolor{PaleGreen4}{rgb}{0.329412,0.545098,0.329412}
  \definecolor{SpringGreen1}{rgb}{0.000000,1.000000,0.498039}
  \definecolor{SpringGreen2}{rgb}{0.000000,0.933333,0.462745}
  \definecolor{SpringGreen3}{rgb}{0.000000,0.803922,0.400000}
  \definecolor{SpringGreen4}{rgb}{0.000000,0.545098,0.270588}
  \definecolor{green1}{rgb}{0.000000,1.000000,0.000000}
  \definecolor{green2}{rgb}{0.000000,0.933333,0.000000}
  \definecolor{green3}{rgb}{0.000000,0.803922,0.000000}
  \definecolor{green4}{rgb}{0.000000,0.545098,0.000000}
  \definecolor{chartreuse1}{rgb}{0.498039,1.000000,0.000000}
  \definecolor{chartreuse2}{rgb}{0.462745,0.933333,0.000000}
  \definecolor{chartreuse3}{rgb}{0.400000,0.803922,0.000000}
  \definecolor{chartreuse4}{rgb}{0.270588,0.545098,0.000000}
  \definecolor{OliveDrab1}{rgb}{0.752941,1.000000,0.243137}
  \definecolor{OliveDrab2}{rgb}{0.701961,0.933333,0.227451}
  \definecolor{OliveDrab3}{rgb}{0.603922,0.803922,0.196078}
  \definecolor{OliveDrab4}{rgb}{0.411765,0.545098,0.133333}
  \definecolor{DarkOliveGreen1}{rgb}{0.792157,1.000000,0.439216}
  \definecolor{DarkOliveGreen2}{rgb}{0.737255,0.933333,0.407843}
  \definecolor{DarkOliveGreen3}{rgb}{0.635294,0.803922,0.352941}
  \definecolor{DarkOliveGreen4}{rgb}{0.431373,0.545098,0.239216}
  \definecolor{khaki1}{rgb}{1.000000,0.964706,0.560784}
  \definecolor{khaki2}{rgb}{0.933333,0.901961,0.521569}
  \definecolor{khaki3}{rgb}{0.803922,0.776471,0.450980}
  \definecolor{khaki4}{rgb}{0.545098,0.525490,0.305882}
  \definecolor{LightGoldenrod1}{rgb}{1.000000,0.925490,0.545098}
  \definecolor{LightGoldenrod2}{rgb}{0.933333,0.862745,0.509804}
  \definecolor{LightGoldenrod3}{rgb}{0.803922,0.745098,0.439216}
  \definecolor{LightGoldenrod4}{rgb}{0.545098,0.505882,0.298039}
  \definecolor{LightYellow1}{rgb}{1.000000,1.000000,0.878431}
  \definecolor{LightYellow2}{rgb}{0.933333,0.933333,0.819608}
  \definecolor{LightYellow3}{rgb}{0.803922,0.803922,0.705882}
  \definecolor{LightYellow4}{rgb}{0.545098,0.545098,0.478431}
  \definecolor{yellow1}{rgb}{1.000000,1.000000,0.000000}
  \definecolor{yellow2}{rgb}{0.933333,0.933333,0.000000}
  \definecolor{yellow3}{rgb}{0.803922,0.803922,0.000000}
  \definecolor{yellow4}{rgb}{0.545098,0.545098,0.000000}
  \definecolor{gold1}{rgb}{1.000000,0.843137,0.000000}
  \definecolor{gold2}{rgb}{0.933333,0.788235,0.000000}
  \definecolor{gold3}{rgb}{0.803922,0.678431,0.000000}
  \definecolor{gold4}{rgb}{0.545098,0.458824,0.000000}
  \definecolor{goldenrod1}{rgb}{1.000000,0.756863,0.145098}
  \definecolor{goldenrod2}{rgb}{0.933333,0.705882,0.133333}
  \definecolor{goldenrod3}{rgb}{0.803922,0.607843,0.113725}
  \definecolor{goldenrod4}{rgb}{0.545098,0.411765,0.078431}
  \definecolor{DarkGoldenrod1}{rgb}{1.000000,0.725490,0.058824}
  \definecolor{DarkGoldenrod2}{rgb}{0.933333,0.678431,0.054902}
  \definecolor{DarkGoldenrod3}{rgb}{0.803922,0.584314,0.047059}
  \definecolor{DarkGoldenrod4}{rgb}{0.545098,0.396078,0.031373}
  \definecolor{RosyBrown1}{rgb}{1.000000,0.756863,0.756863}
  \definecolor{RosyBrown2}{rgb}{0.933333,0.705882,0.705882}
  \definecolor{RosyBrown3}{rgb}{0.803922,0.607843,0.607843}
  \definecolor{RosyBrown4}{rgb}{0.545098,0.411765,0.411765}
  \definecolor{IndianRed1}{rgb}{1.000000,0.415686,0.415686}
  \definecolor{IndianRed2}{rgb}{0.933333,0.388235,0.388235}
  \definecolor{IndianRed3}{rgb}{0.803922,0.333333,0.333333}
  \definecolor{IndianRed4}{rgb}{0.545098,0.227451,0.227451}
  \definecolor{sienna1}{rgb}{1.000000,0.509804,0.278431}
  \definecolor{sienna2}{rgb}{0.933333,0.474510,0.258824}
  \definecolor{sienna3}{rgb}{0.803922,0.407843,0.223529}
  \definecolor{sienna4}{rgb}{0.545098,0.278431,0.149020}
  \definecolor{burlywood1}{rgb}{1.000000,0.827451,0.607843}
  \definecolor{burlywood2}{rgb}{0.933333,0.772549,0.568627}
  \definecolor{burlywood3}{rgb}{0.803922,0.666667,0.490196}
  \definecolor{burlywood4}{rgb}{0.545098,0.450980,0.333333}
  \definecolor{wheat1}{rgb}{1.000000,0.905882,0.729412}
  \definecolor{wheat2}{rgb}{0.933333,0.847059,0.682353}
  \definecolor{wheat3}{rgb}{0.803922,0.729412,0.588235}
  \definecolor{wheat4}{rgb}{0.545098,0.494118,0.400000}
  \definecolor{tan1}{rgb}{1.000000,0.647059,0.309804}
  \definecolor{tan2}{rgb}{0.933333,0.603922,0.286275}
  \definecolor{tan3}{rgb}{0.803922,0.521569,0.247059}
  \definecolor{tan4}{rgb}{0.545098,0.352941,0.168627}
  \definecolor{chocolate1}{rgb}{1.000000,0.498039,0.141176}
  \definecolor{chocolate2}{rgb}{0.933333,0.462745,0.129412}
  \definecolor{chocolate3}{rgb}{0.803922,0.400000,0.113725}
  \definecolor{chocolate4}{rgb}{0.545098,0.270588,0.074510}
  \definecolor{firebrick1}{rgb}{1.000000,0.188235,0.188235}
  \definecolor{firebrick2}{rgb}{0.933333,0.172549,0.172549}
  \definecolor{firebrick3}{rgb}{0.803922,0.149020,0.149020}
  \definecolor{firebrick4}{rgb}{0.545098,0.101961,0.101961}
  \definecolor{brown1}{rgb}{1.000000,0.250980,0.250980}
  \definecolor{brown2}{rgb}{0.933333,0.231373,0.231373}
  \definecolor{brown3}{rgb}{0.803922,0.200000,0.200000}
  \definecolor{brown4}{rgb}{0.545098,0.137255,0.137255}
  \definecolor{salmon1}{rgb}{1.000000,0.549020,0.411765}
  \definecolor{salmon2}{rgb}{0.933333,0.509804,0.384314}
  \definecolor{salmon3}{rgb}{0.803922,0.439216,0.329412}
  \definecolor{salmon4}{rgb}{0.545098,0.298039,0.223529}
  \definecolor{LightSalmon1}{rgb}{1.000000,0.627451,0.478431}
  \definecolor{LightSalmon2}{rgb}{0.933333,0.584314,0.447059}
  \definecolor{LightSalmon3}{rgb}{0.803922,0.505882,0.384314}
  \definecolor{LightSalmon4}{rgb}{0.545098,0.341176,0.258824}
  \definecolor{orange1}{rgb}{1.000000,0.647059,0.000000}
  \definecolor{orange2}{rgb}{0.933333,0.603922,0.000000}
  \definecolor{orange3}{rgb}{0.803922,0.521569,0.000000}
  \definecolor{orange4}{rgb}{0.545098,0.352941,0.000000}
  \definecolor{DarkOrange1}{rgb}{1.000000,0.498039,0.000000}
  \definecolor{DarkOrange2}{rgb}{0.933333,0.462745,0.000000}
  \definecolor{DarkOrange3}{rgb}{0.803922,0.400000,0.000000}
  \definecolor{DarkOrange4}{rgb}{0.545098,0.270588,0.000000}
  \definecolor{coral1}{rgb}{1.000000,0.447059,0.337255}
  \definecolor{coral2}{rgb}{0.933333,0.415686,0.313726}
  \definecolor{coral3}{rgb}{0.803922,0.356863,0.270588}
  \definecolor{coral4}{rgb}{0.545098,0.243137,0.184314}
  \definecolor{tomato1}{rgb}{1.000000,0.388235,0.278431}
  \definecolor{tomato2}{rgb}{0.933333,0.360784,0.258824}
  \definecolor{tomato3}{rgb}{0.803922,0.309804,0.223529}
  \definecolor{tomato4}{rgb}{0.545098,0.211765,0.149020}
  \definecolor{OrangeRed1}{rgb}{1.000000,0.270588,0.000000}
  \definecolor{OrangeRed2}{rgb}{0.933333,0.250980,0.000000}
  \definecolor{OrangeRed3}{rgb}{0.803922,0.215686,0.000000}
  \definecolor{OrangeRed4}{rgb}{0.545098,0.145098,0.000000}
  \definecolor{red1}{rgb}{1.000000,0.000000,0.000000}
  \definecolor{red2}{rgb}{0.933333,0.000000,0.000000}
  \definecolor{red3}{rgb}{0.803922,0.000000,0.000000}
  \definecolor{red4}{rgb}{0.545098,0.000000,0.000000}
  \definecolor{DeepPink1}{rgb}{1.000000,0.078431,0.576471}
  \definecolor{DeepPink2}{rgb}{0.933333,0.070588,0.537255}
  \definecolor{DeepPink3}{rgb}{0.803922,0.062745,0.462745}
  \definecolor{DeepPink4}{rgb}{0.545098,0.039216,0.313726}
  \definecolor{HotPink1}{rgb}{1.000000,0.431373,0.705882}
  \definecolor{HotPink2}{rgb}{0.933333,0.415686,0.654902}
  \definecolor{HotPink3}{rgb}{0.803922,0.376471,0.564706}
  \definecolor{HotPink4}{rgb}{0.545098,0.227451,0.384314}
  \definecolor{pink1}{rgb}{1.000000,0.709804,0.772549}
  \definecolor{pink2}{rgb}{0.933333,0.662745,0.721569}
  \definecolor{pink3}{rgb}{0.803922,0.568627,0.619608}
  \definecolor{pink4}{rgb}{0.545098,0.388235,0.423529}
  \definecolor{LightPink1}{rgb}{1.000000,0.682353,0.725490}
  \definecolor{LightPink2}{rgb}{0.933333,0.635294,0.678431}
  \definecolor{LightPink3}{rgb}{0.803922,0.549020,0.584314}
  \definecolor{LightPink4}{rgb}{0.545098,0.372549,0.396078}
  \definecolor{PaleVioletRed1}{rgb}{1.000000,0.509804,0.670588}
  \definecolor{PaleVioletRed2}{rgb}{0.933333,0.474510,0.623529}
  \definecolor{PaleVioletRed3}{rgb}{0.803922,0.407843,0.537255}
  \definecolor{PaleVioletRed4}{rgb}{0.545098,0.278431,0.364706}
  \definecolor{maroon1}{rgb}{1.000000,0.203922,0.701961}
  \definecolor{maroon2}{rgb}{0.933333,0.188235,0.654902}
  \definecolor{maroon3}{rgb}{0.803922,0.160784,0.564706}
  \definecolor{maroon4}{rgb}{0.545098,0.109804,0.384314}
  \definecolor{VioletRed1}{rgb}{1.000000,0.243137,0.588235}
  \definecolor{VioletRed2}{rgb}{0.933333,0.227451,0.549020}
  \definecolor{VioletRed3}{rgb}{0.803922,0.196078,0.470588}
  \definecolor{VioletRed4}{rgb}{0.545098,0.133333,0.321569}
  \definecolor{magenta1}{rgb}{1.000000,0.000000,1.000000}
  \definecolor{magenta2}{rgb}{0.933333,0.000000,0.933333}
  \definecolor{magenta3}{rgb}{0.803922,0.000000,0.803922}
  \definecolor{magenta4}{rgb}{0.545098,0.000000,0.545098}
  \definecolor{orchid1}{rgb}{1.000000,0.513726,0.980392}
  \definecolor{orchid2}{rgb}{0.933333,0.478431,0.913725}
  \definecolor{orchid3}{rgb}{0.803922,0.411765,0.788235}
  \definecolor{orchid4}{rgb}{0.545098,0.278431,0.537255}
  \definecolor{plum1}{rgb}{1.000000,0.733333,1.000000}
  \definecolor{plum2}{rgb}{0.933333,0.682353,0.933333}
  \definecolor{plum3}{rgb}{0.803922,0.588235,0.803922}
  \definecolor{plum4}{rgb}{0.545098,0.400000,0.545098}
  \definecolor{MediumOrchid1}{rgb}{0.878431,0.400000,1.000000}
  \definecolor{MediumOrchid2}{rgb}{0.819608,0.372549,0.933333}
  \definecolor{MediumOrchid3}{rgb}{0.705882,0.321569,0.803922}
  \definecolor{MediumOrchid4}{rgb}{0.478431,0.215686,0.545098}
  \definecolor{DarkOrchid1}{rgb}{0.749020,0.243137,1.000000}
  \definecolor{DarkOrchid2}{rgb}{0.698039,0.227451,0.933333}
  \definecolor{DarkOrchid3}{rgb}{0.603922,0.196078,0.803922}
  \definecolor{DarkOrchid4}{rgb}{0.407843,0.133333,0.545098}
  \definecolor{purple1}{rgb}{0.607843,0.188235,1.000000}
  \definecolor{purple2}{rgb}{0.568627,0.172549,0.933333}
  \definecolor{purple3}{rgb}{0.490196,0.149020,0.803922}
  \definecolor{purple4}{rgb}{0.333333,0.101961,0.545098}
  \definecolor{MediumPurple1}{rgb}{0.670588,0.509804,1.000000}
  \definecolor{MediumPurple2}{rgb}{0.623529,0.474510,0.933333}
  \definecolor{MediumPurple3}{rgb}{0.537255,0.407843,0.803922}
  \definecolor{MediumPurple4}{rgb}{0.364706,0.278431,0.545098}
  \definecolor{thistle1}{rgb}{1.000000,0.882353,1.000000}
  \definecolor{thistle2}{rgb}{0.933333,0.823529,0.933333}
  \definecolor{thistle3}{rgb}{0.803922,0.709804,0.803922}
  \definecolor{thistle4}{rgb}{0.545098,0.482353,0.545098}
  \definecolor{gray0}{rgb}{0.000000,0.000000,0.000000}
  \definecolor{grey0}{rgb}{0.000000,0.000000,0.000000}
  \definecolor{gray1}{rgb}{0.011765,0.011765,0.011765}
  \definecolor{grey1}{rgb}{0.011765,0.011765,0.011765}
  \definecolor{gray2}{rgb}{0.019608,0.019608,0.019608}
  \definecolor{grey2}{rgb}{0.019608,0.019608,0.019608}
  \definecolor{gray3}{rgb}{0.031373,0.031373,0.031373}
  \definecolor{grey3}{rgb}{0.031373,0.031373,0.031373}
  \definecolor{gray4}{rgb}{0.039216,0.039216,0.039216}
  \definecolor{grey4}{rgb}{0.039216,0.039216,0.039216}
  \definecolor{gray5}{rgb}{0.050980,0.050980,0.050980}
  \definecolor{grey5}{rgb}{0.050980,0.050980,0.050980}
  \definecolor{gray6}{rgb}{0.058824,0.058824,0.058824}
  \definecolor{grey6}{rgb}{0.058824,0.058824,0.058824}
  \definecolor{gray7}{rgb}{0.070588,0.070588,0.070588}
  \definecolor{grey7}{rgb}{0.070588,0.070588,0.070588}
  \definecolor{gray8}{rgb}{0.078431,0.078431,0.078431}
  \definecolor{grey8}{rgb}{0.078431,0.078431,0.078431}
  \definecolor{gray9}{rgb}{0.090196,0.090196,0.090196}
  \definecolor{grey9}{rgb}{0.090196,0.090196,0.090196}
  \definecolor{gray10}{rgb}{0.101961,0.101961,0.101961}
  \definecolor{grey10}{rgb}{0.101961,0.101961,0.101961}
  \definecolor{gray11}{rgb}{0.109804,0.109804,0.109804}
  \definecolor{grey11}{rgb}{0.109804,0.109804,0.109804}
  \definecolor{gray12}{rgb}{0.121569,0.121569,0.121569}
  \definecolor{grey12}{rgb}{0.121569,0.121569,0.121569}
  \definecolor{gray13}{rgb}{0.129412,0.129412,0.129412}
  \definecolor{grey13}{rgb}{0.129412,0.129412,0.129412}
  \definecolor{gray14}{rgb}{0.141176,0.141176,0.141176}
  \definecolor{grey14}{rgb}{0.141176,0.141176,0.141176}
  \definecolor{gray15}{rgb}{0.149020,0.149020,0.149020}
  \definecolor{grey15}{rgb}{0.149020,0.149020,0.149020}
  \definecolor{gray16}{rgb}{0.160784,0.160784,0.160784}
  \definecolor{grey16}{rgb}{0.160784,0.160784,0.160784}
  \definecolor{gray17}{rgb}{0.168627,0.168627,0.168627}
  \definecolor{grey17}{rgb}{0.168627,0.168627,0.168627}
  \definecolor{gray18}{rgb}{0.180392,0.180392,0.180392}
  \definecolor{grey18}{rgb}{0.180392,0.180392,0.180392}
  \definecolor{gray19}{rgb}{0.188235,0.188235,0.188235}
  \definecolor{grey19}{rgb}{0.188235,0.188235,0.188235}
  \definecolor{gray20}{rgb}{0.200000,0.200000,0.200000}
  \definecolor{grey20}{rgb}{0.200000,0.200000,0.200000}
  \definecolor{gray21}{rgb}{0.211765,0.211765,0.211765}
  \definecolor{grey21}{rgb}{0.211765,0.211765,0.211765}
  \definecolor{gray22}{rgb}{0.219608,0.219608,0.219608}
  \definecolor{grey22}{rgb}{0.219608,0.219608,0.219608}
  \definecolor{gray23}{rgb}{0.231373,0.231373,0.231373}
  \definecolor{grey23}{rgb}{0.231373,0.231373,0.231373}
  \definecolor{gray24}{rgb}{0.239216,0.239216,0.239216}
  \definecolor{grey24}{rgb}{0.239216,0.239216,0.239216}
  \definecolor{gray25}{rgb}{0.250980,0.250980,0.250980}
  \definecolor{grey25}{rgb}{0.250980,0.250980,0.250980}
  \definecolor{gray26}{rgb}{0.258824,0.258824,0.258824}
  \definecolor{grey26}{rgb}{0.258824,0.258824,0.258824}
  \definecolor{gray27}{rgb}{0.270588,0.270588,0.270588}
  \definecolor{grey27}{rgb}{0.270588,0.270588,0.270588}
  \definecolor{gray28}{rgb}{0.278431,0.278431,0.278431}
  \definecolor{grey28}{rgb}{0.278431,0.278431,0.278431}
  \definecolor{gray29}{rgb}{0.290196,0.290196,0.290196}
  \definecolor{grey29}{rgb}{0.290196,0.290196,0.290196}
  \definecolor{gray30}{rgb}{0.301961,0.301961,0.301961}
  \definecolor{grey30}{rgb}{0.301961,0.301961,0.301961}
  \definecolor{gray31}{rgb}{0.309804,0.309804,0.309804}
  \definecolor{grey31}{rgb}{0.309804,0.309804,0.309804}
  \definecolor{gray32}{rgb}{0.321569,0.321569,0.321569}
  \definecolor{grey32}{rgb}{0.321569,0.321569,0.321569}
  \definecolor{gray33}{rgb}{0.329412,0.329412,0.329412}
  \definecolor{grey33}{rgb}{0.329412,0.329412,0.329412}
  \definecolor{gray34}{rgb}{0.341176,0.341176,0.341176}
  \definecolor{grey34}{rgb}{0.341176,0.341176,0.341176}
  \definecolor{gray35}{rgb}{0.349020,0.349020,0.349020}
  \definecolor{grey35}{rgb}{0.349020,0.349020,0.349020}
  \definecolor{gray36}{rgb}{0.360784,0.360784,0.360784}
  \definecolor{grey36}{rgb}{0.360784,0.360784,0.360784}
  \definecolor{gray37}{rgb}{0.368627,0.368627,0.368627}
  \definecolor{grey37}{rgb}{0.368627,0.368627,0.368627}
  \definecolor{gray38}{rgb}{0.380392,0.380392,0.380392}
  \definecolor{grey38}{rgb}{0.380392,0.380392,0.380392}
  \definecolor{gray39}{rgb}{0.388235,0.388235,0.388235}
  \definecolor{grey39}{rgb}{0.388235,0.388235,0.388235}
  \definecolor{gray40}{rgb}{0.400000,0.400000,0.400000}
  \definecolor{grey40}{rgb}{0.400000,0.400000,0.400000}
  \definecolor{gray41}{rgb}{0.411765,0.411765,0.411765}
  \definecolor{grey41}{rgb}{0.411765,0.411765,0.411765}
  \definecolor{gray42}{rgb}{0.419608,0.419608,0.419608}
  \definecolor{grey42}{rgb}{0.419608,0.419608,0.419608}
  \definecolor{gray43}{rgb}{0.431373,0.431373,0.431373}
  \definecolor{grey43}{rgb}{0.431373,0.431373,0.431373}
  \definecolor{gray44}{rgb}{0.439216,0.439216,0.439216}
  \definecolor{grey44}{rgb}{0.439216,0.439216,0.439216}
  \definecolor{gray45}{rgb}{0.450980,0.450980,0.450980}
  \definecolor{grey45}{rgb}{0.450980,0.450980,0.450980}
  \definecolor{gray46}{rgb}{0.458824,0.458824,0.458824}
  \definecolor{grey46}{rgb}{0.458824,0.458824,0.458824}
  \definecolor{gray47}{rgb}{0.470588,0.470588,0.470588}
  \definecolor{grey47}{rgb}{0.470588,0.470588,0.470588}
  \definecolor{gray48}{rgb}{0.478431,0.478431,0.478431}
  \definecolor{grey48}{rgb}{0.478431,0.478431,0.478431}
  \definecolor{gray49}{rgb}{0.490196,0.490196,0.490196}
  \definecolor{grey49}{rgb}{0.490196,0.490196,0.490196}
  \definecolor{gray50}{rgb}{0.498039,0.498039,0.498039}
  \definecolor{grey50}{rgb}{0.498039,0.498039,0.498039}
  \definecolor{gray51}{rgb}{0.509804,0.509804,0.509804}
  \definecolor{grey51}{rgb}{0.509804,0.509804,0.509804}
  \definecolor{gray52}{rgb}{0.521569,0.521569,0.521569}
  \definecolor{grey52}{rgb}{0.521569,0.521569,0.521569}
  \definecolor{gray53}{rgb}{0.529412,0.529412,0.529412}
  \definecolor{grey53}{rgb}{0.529412,0.529412,0.529412}
  \definecolor{gray54}{rgb}{0.541176,0.541176,0.541176}
  \definecolor{grey54}{rgb}{0.541176,0.541176,0.541176}
  \definecolor{gray55}{rgb}{0.549020,0.549020,0.549020}
  \definecolor{grey55}{rgb}{0.549020,0.549020,0.549020}
  \definecolor{gray56}{rgb}{0.560784,0.560784,0.560784}
  \definecolor{grey56}{rgb}{0.560784,0.560784,0.560784}
  \definecolor{gray57}{rgb}{0.568627,0.568627,0.568627}
  \definecolor{grey57}{rgb}{0.568627,0.568627,0.568627}
  \definecolor{gray58}{rgb}{0.580392,0.580392,0.580392}
  \definecolor{grey58}{rgb}{0.580392,0.580392,0.580392}
  \definecolor{gray59}{rgb}{0.588235,0.588235,0.588235}
  \definecolor{grey59}{rgb}{0.588235,0.588235,0.588235}
  \definecolor{gray60}{rgb}{0.600000,0.600000,0.600000}
  \definecolor{grey60}{rgb}{0.600000,0.600000,0.600000}
  \definecolor{gray61}{rgb}{0.611765,0.611765,0.611765}
  \definecolor{grey61}{rgb}{0.611765,0.611765,0.611765}
  \definecolor{gray62}{rgb}{0.619608,0.619608,0.619608}
  \definecolor{grey62}{rgb}{0.619608,0.619608,0.619608}
  \definecolor{gray63}{rgb}{0.631373,0.631373,0.631373}
  \definecolor{grey63}{rgb}{0.631373,0.631373,0.631373}
  \definecolor{gray64}{rgb}{0.639216,0.639216,0.639216}
  \definecolor{grey64}{rgb}{0.639216,0.639216,0.639216}
  \definecolor{gray65}{rgb}{0.650980,0.650980,0.650980}
  \definecolor{grey65}{rgb}{0.650980,0.650980,0.650980}
  \definecolor{gray66}{rgb}{0.658824,0.658824,0.658824}
  \definecolor{grey66}{rgb}{0.658824,0.658824,0.658824}
  \definecolor{gray67}{rgb}{0.670588,0.670588,0.670588}
  \definecolor{grey67}{rgb}{0.670588,0.670588,0.670588}
  \definecolor{gray68}{rgb}{0.678431,0.678431,0.678431}
  \definecolor{grey68}{rgb}{0.678431,0.678431,0.678431}
  \definecolor{gray69}{rgb}{0.690196,0.690196,0.690196}
  \definecolor{grey69}{rgb}{0.690196,0.690196,0.690196}
  \definecolor{gray70}{rgb}{0.701961,0.701961,0.701961}
  \definecolor{grey70}{rgb}{0.701961,0.701961,0.701961}
  \definecolor{gray71}{rgb}{0.709804,0.709804,0.709804}
  \definecolor{grey71}{rgb}{0.709804,0.709804,0.709804}
  \definecolor{gray72}{rgb}{0.721569,0.721569,0.721569}
  \definecolor{grey72}{rgb}{0.721569,0.721569,0.721569}
  \definecolor{gray73}{rgb}{0.729412,0.729412,0.729412}
  \definecolor{grey73}{rgb}{0.729412,0.729412,0.729412}
  \definecolor{gray74}{rgb}{0.741176,0.741176,0.741176}
  \definecolor{grey74}{rgb}{0.741176,0.741176,0.741176}
  \definecolor{gray75}{rgb}{0.749020,0.749020,0.749020}
  \definecolor{grey75}{rgb}{0.749020,0.749020,0.749020}
  \definecolor{gray76}{rgb}{0.760784,0.760784,0.760784}
  \definecolor{grey76}{rgb}{0.760784,0.760784,0.760784}
  \definecolor{gray77}{rgb}{0.768627,0.768627,0.768627}
  \definecolor{grey77}{rgb}{0.768627,0.768627,0.768627}
  \definecolor{gray78}{rgb}{0.780392,0.780392,0.780392}
  \definecolor{grey78}{rgb}{0.780392,0.780392,0.780392}
  \definecolor{gray79}{rgb}{0.788235,0.788235,0.788235}
  \definecolor{grey79}{rgb}{0.788235,0.788235,0.788235}
  \definecolor{gray80}{rgb}{0.800000,0.800000,0.800000}
  \definecolor{grey80}{rgb}{0.800000,0.800000,0.800000}
  \definecolor{gray81}{rgb}{0.811765,0.811765,0.811765}
  \definecolor{grey81}{rgb}{0.811765,0.811765,0.811765}
  \definecolor{gray82}{rgb}{0.819608,0.819608,0.819608}
  \definecolor{grey82}{rgb}{0.819608,0.819608,0.819608}
  \definecolor{gray83}{rgb}{0.831373,0.831373,0.831373}
  \definecolor{grey83}{rgb}{0.831373,0.831373,0.831373}
  \definecolor{gray84}{rgb}{0.839216,0.839216,0.839216}
  \definecolor{grey84}{rgb}{0.839216,0.839216,0.839216}
  \definecolor{gray85}{rgb}{0.850980,0.850980,0.850980}
  \definecolor{grey85}{rgb}{0.850980,0.850980,0.850980}
  \definecolor{gray86}{rgb}{0.858824,0.858824,0.858824}
  \definecolor{grey86}{rgb}{0.858824,0.858824,0.858824}
  \definecolor{gray87}{rgb}{0.870588,0.870588,0.870588}
  \definecolor{grey87}{rgb}{0.870588,0.870588,0.870588}
  \definecolor{gray88}{rgb}{0.878431,0.878431,0.878431}
  \definecolor{grey88}{rgb}{0.878431,0.878431,0.878431}
  \definecolor{gray89}{rgb}{0.890196,0.890196,0.890196}
  \definecolor{grey89}{rgb}{0.890196,0.890196,0.890196}
  \definecolor{gray90}{rgb}{0.898039,0.898039,0.898039}
  \definecolor{grey90}{rgb}{0.898039,0.898039,0.898039}
  \definecolor{gray91}{rgb}{0.909804,0.909804,0.909804}
  \definecolor{grey91}{rgb}{0.909804,0.909804,0.909804}
  \definecolor{gray92}{rgb}{0.921569,0.921569,0.921569}
  \definecolor{grey92}{rgb}{0.921569,0.921569,0.921569}
  \definecolor{gray93}{rgb}{0.929412,0.929412,0.929412}
  \definecolor{grey93}{rgb}{0.929412,0.929412,0.929412}
  \definecolor{gray94}{rgb}{0.941176,0.941176,0.941176}
  \definecolor{grey94}{rgb}{0.941176,0.941176,0.941176}
  \definecolor{gray95}{rgb}{0.949020,0.949020,0.949020}
  \definecolor{grey95}{rgb}{0.949020,0.949020,0.949020}
  \definecolor{gray96}{rgb}{0.960784,0.960784,0.960784}
  \definecolor{grey96}{rgb}{0.960784,0.960784,0.960784}
  \definecolor{gray97}{rgb}{0.968627,0.968627,0.968627}
  \definecolor{grey97}{rgb}{0.968627,0.968627,0.968627}
  \definecolor{gray98}{rgb}{0.980392,0.980392,0.980392}
  \definecolor{grey98}{rgb}{0.980392,0.980392,0.980392}
  \definecolor{gray99}{rgb}{0.988235,0.988235,0.988235}
  \definecolor{grey99}{rgb}{0.988235,0.988235,0.988235}
  \definecolor{gray100}{rgb}{1.000000,1.000000,1.000000}
  \definecolor{grey100}{rgb}{1.000000,1.000000,1.000000}
  \definecolor{dark grey}{rgb}{0.662745,0.662745,0.662745}
  \definecolor{DarkGrey}{rgb}{0.662745,0.662745,0.662745}
  \definecolor{dark gray}{rgb}{0.662745,0.662745,0.662745}
  \definecolor{DarkGray}{rgb}{0.662745,0.662745,0.662745}
  \definecolor{dark blue}{rgb}{0.000000,0.000000,0.545098}
  \definecolor{DarkBlue}{rgb}{0.000000,0.000000,0.545098}
  \definecolor{dark cyan}{rgb}{0.000000,0.545098,0.545098}
  \definecolor{DarkCyan}{rgb}{0.000000,0.545098,0.545098}
  \definecolor{dark magenta}{rgb}{0.545098,0.000000,0.545098}
  \definecolor{DarkMagenta}{rgb}{0.545098,0.000000,0.545098}
  \definecolor{dark red}{rgb}{0.545098,0.000000,0.000000}
  \definecolor{DarkRed}{rgb}{0.545098,0.000000,0.000000}
  \definecolor{light green}{rgb}{0.564706,0.933333,0.564706}
  \definecolor{LightGreen}{rgb}{0.564706,0.933333,0.564706}

%% file: SOURCES/abstract-thermo-AMSA.tex

\noindent
{\bf\abstractname{.}}
In this paper we apply the entropy principle to the relativistic
version of the differential equations describing 
a standard fluid flow, that is, the equations for mass, momentum,
and a system for the energy matrix.
These are the second order equations
which have been introduced in \CIT{Alt2016}{}.
Since the principle also says that the entropy equation is 
a scalar equation,
this implies, as we show, that one has to take
a trace in the energy part of the system.
Thus one arrives at the relativistic mass-momentum-energy system 
for the fluid.
In the procedure we use the well-known Liu-M\"uller sum \CIT{Mueller1985}{}
in order to deduce the Gibbs relation and the residual entropy inequality.

\vspace*{2cm}
\noindent
{\bf Version:}
Some minor misprints corrected and unessential improvements.

%% file: SOURCES/content.tex
\input{SOURCES/01intro}


\newpage
\input{SOURCES/02general}
\input{SOURCES/03lagrange}
\input{SOURCES/04theorem}
\input{SOURCES/05exploit}
\input{SOURCES/06relativ}

\newpage
\input{SOURCES/07constit}

%% file: SOURCES/01intro.tex
\sect{intro}{\DE{Einf\"uhrung}\EN{Introduction}}
It has been a long history for the entropy principle to come up
to the essential dif\-fe\-ren\-tial inequality
\begin{equ}{entropy}
  \sigma:=\de_t\eta+\div_x\psi\geq0
\end{equ}
in classical coordinates $(t,x)$,
$x=(x_1\til x_n)$ where $n=3$ is the physical case.
Here $\eta$ is the entropy and $\psi$ the entropy flux,
hence $\tx\eta=(\eta,\psi)$ are the total entropy quantities.
This principle has been successfully applied to the mass-momentum-energy system
in many physical examples. 
The history started $\approx150$ years ago and
one can find this principle in many books, among them 
Prigogine \CIT{Prigogine1954}{Chapter III}, 
DeGroot \& Mazur \CIT{DeGrootMazur}{Chapter III}, 
Truesdell \& Noll \CIT{TruesdellNoll}{D.II}, 
Truesdell \CIT{Truesdell1969}{}, 
Ingo M\"uller \CIT{Mueller1973}{Kapitel IV}, 
to mention a few, 
which were all published in the period 1954-1973.
It is part of the entropy principle 
that the differential equation $\sigma=\de_t\eta+\div_x\psi$
is an objective scalar equation, see \CIT{Alt-Kontinuum}{Sec II.3},
by which we mean that for 
the weak equation 
\[
  \int(\de_t\zeta\cdot\eta+\grad\zeta\dd\psi+\zeta\cdot\sigma)\d\leb{n+1}=0
\]
the test function $\zeta$ is an objective scalar, that is
$\zeta\circ Y=\zeta^*$, where $Y$ is the observer transformation.
This is satisfied, see \CIT{Alt-Kontinuum}{Sec I.5},
if $\eta$ is an objective scalar, that is $\eta\circ Y=\eta^*$,
and $\psi$ satisfies $\psi\circ Y=\eta^*\dot{X}+Q\psi^*$.

\medskip
In the relativistic case one formulates 
the entropy principle in the form
\begin{equ}{entropy-relativ}
  \sigma:=\sum_{j\geq0}\de_{y_j}\tx\eta\geq0
\end{equ}
with 4-dimensional coordinates $y=(y_0,y_1\til y_n)$,
again $n=3$ in the physical case.
As postulate we assume that the weak version
\begin{Equ}{weak-relativ}
  \int\big(\sum_{j\geq0}\de_{y_j}\zeta\cdot\tx\eta
  +\zeta\cdot\sigma\big)\d\leb{n+1}=0
\end{Equ}
is satisfied for objective test functions $\zeta$,
that is $\zeta\circ Y=\zeta^*$.
Here $Y$ is a re\-la\-ti\-vis\-tic ob\-ser\-ver transformation.
This is satisfied,
see \CIT{Alt-Kontinuum}{Sec I.5},
if the 4-entropy vector $\tx\eta$
satisfies $\tx\eta\circ Y=\D{Y}\tx\eta^*$,
that is, $\tx\eta$ is a contravariant vector
(see the definition below). 

\medskip
The re\-la\-ti\-vistic case one finds also in sections of the books of
Ingo M\"uller \CIT{Mueller1985}{} and
M\"uller \& Ruggeri \CIT{Mueller1998}{}.
%
Here we take advantage of this principle \EQU{entropy-relativ} and apply
it to the relativistic system \CIT{Alt2016}{(10.2)} 
\begin{Equ}{sys}
  \sum_{j\geq0}\de_{y_j}T_{\alpha j}=g_\alpha
  \quad\tm{for} \alpha\in\{0\til n\}^N
\end{Equ}
which we have developed in \CIT{Alt2016}{}. 
But here we will take it only 
for $N=2$, that is, we write $\alpha=(k,l)$ with $k,l\geq0$,
and we use a representation which is made for gases and fluids
\begin{Equ}{structure}
  T_{klj}=(\rho\tx v_k\tx v_l + E_{kl})\tx v_j + \tilde{\rm Q}_{klj}
\end{Equ}
where $\tx v$ is the four dimensional fluid velocity,
see the definition in \CIT{Alt2016}{5.2},
and with the assumptions \EQU{theorem.*.0} on $E$ and $\tilde{\rm Q}$.
It should be said that the right-hand side $g_{kl}$ of this system
contains the Coriolis coefficients
and of course external or internal forces. 

\medskip
Altogether,
this system includes the mass-momentum system and a system describing
the energy matrix $E$.
The entropy principle
for gases and fluids, see section \REF{exploit..} and \REF{relativ..},
forces us to perform a trace of the energy matrix equation
in order to have an entropy $\eta$ which is an objective scalar.  
This method is even new for the classical fluid case.
You will find the result 
in the final theorem~s\REF{theorem.entropy.}.
It contains the statement that the residual inequality $\sigma\geq0$,
that is, 
the entropy production \EQU{theorem.entropy.sigma} is non-negative.
Also as a consequence of the entropy principle
there are some important identities.
So the entropy $\eta=\hat\eta(\rho,\eps)$
is a function of the density $\rho$ and the internal energy
\[
  \eps=\frac12(\up{{\rm P}}T\up\GG{-1}{\rm P})\ddd E
\,.\]
And the system \EQU{sys} is specified by two equations,
the mass-momentum and the energy
equation, see \REF{theorem.reduced.mm} and \REF{theorem.H.},
\[\begin{arr}c
  \sum_{j\geq0}\de_{y_j}\big(
  \rho\tx v_k\tx v_j + \tx v_k\tx\JJ_j + \tx\Pi_{kj}
  \big)=g_{k}
\,,\\
  \sum_{j\geq0}\de_{y_j}\Big(
  \big(\frac\rho2\,
  \tx v\dd(\up{{\rm P}}T\up\GG{-1}{\rm P})\tx v+\eps\big)\,\tx v_j
  + \tilde q_j
  \Big)=g
\,,\end{arr}\]
with
\[\begin{arr}c
  \tx\Pi := p\,({\rm P}\GG\up{{\rm P}}T) - \tx S
\,,\\
  \tilde q:=\frac\rho2\,(\tx v\dd(\up{{\rm P}}T\up\GG{-1}{\rm P})\tx v)\,\JJ
  +\tx v(\up{{\rm P}}T\up\GG{-1}{\rm P})\tx\Pi
  +\tx q
\,,\end{arr}\]
where the inequality restrictions are in the residual inequality $\sigma\geq0$,
see \REF{theorem.entropy.} for details about the entropy production $\sigma$
and the total entropy $\tx\eta$.

\medskip
Both, the mass-momentum system and the energy equation are reductions
of the equations we started with.
The statement \REF{theorem.entropy.}
is the entropy principle in the simplest case.
More complicated versions one expects
in the case that $\eta$ might, for example, depend on gradients
as in the classical case, or on other vectorial quantities,
because the whole system then is more complicated.

\PAR
\PAR
\begin{rem}{Notation}
The definition of a contravariant $M$-tensor
$T=\seq{T_{k_1\cdots k_M}}{k_1\til k_M}$ is 
\begin{Equ}{contra}
   T_{k_1\cdots k_M}\circ Y=\sum_{\bar k_1\til\bar k_M\geq0}
  Y_{k_1\p\bar k_1}\cdots Y_{k_M\p\bar k_M}T^*_{\bar k_1\cdots\bar k_M}
\,,\end{Equ}
and the definition of a covariant $M$-tensor
$T=\seq{T_{k_1\cdots k_M}}{k_1\til k_M}$ 
\begin{Equ}{co}
  T^*_{\bar k_1\cdots\bar k_M}=\sum_{k_1\til k_M\geq0}
  Y_{k_1\p\bar k_1}\cdots Y_{k_M\p\bar k_M} T_{k_1\cdots k_M}\circ Y
\,.\end{Equ}
Here $y=Y(y^*)$ is the observer transformation.
\end{rem}

%% file: SOURCES/02general.tex
\sect{general}{\DE{Allgemeine Momente}\EN{General moments}}
The version of moments of order less or equal $N$ is
\begin{Equ}{sys}
  \sum_{j\geq0}\de_{y_j}T_{\alpha j}=g_\alpha
  \quad\tm{\DE{f\"ur}\EN{for}} \alpha\in\{0\til n\}^N
\end{Equ}
with a tensor $T=\seq{T_\beta}{\beta\in\{0\til n\}^{N+1}}$
which has to be symmetric only in the first $N$ components
of the multiindex $\beta=(\beta_1\til\beta_M)$, $M:=N+1$,
that is, setting $\beta=(\alpha,j)$ as in the equations,
$T_{\alpha j}$ and $g_\alpha$ are symmetric in the components of $\alpha$.
Here $y\in\UU\subset\RR^{n+1}$ where
$y=(y_0\til y_n)$ and $n=3$ in the physical situation.
See \CIT{Alt2016}{10 Higher moments} for more information.
System \EQU{sys} is equivalent to the weak version
\begin{Equ}{sys-weak}
  \sum_\alpha\int_\UU\Big(
  \sum_{j\geq0}\de_{y_j}\zeta_\alpha\cdot T_{\alpha j}+\zeta_\alpha g_\alpha
  \Big)=0
  \quad\tm{\DE{f\"ur}\EN{for}} \zeta_\alpha\in C_0^\infty(\UU)
\,,\end{Equ}
where the physical type of the system  
is defined by the fact that the test function
$\zeta:=\seq{\zeta_\alpha}{\alpha}$ is a covariant $N$-tensor,
that is it satisfies the transformation rule
\begin{Equ}{rule-zeta}
  \zeta^*_{\bar\alpha}
  =\sum_{\alpha}
  Y_{\alpha_1\p\bar\alpha_1}\cdots Y_{\alpha_N\p\bar\alpha_N}
  \zeta_\alpha\circ Y
\,.\end{Equ}
This is satisfied, see \CIT{Alt-Kontinuum}{Chap I {\S}5},
if $T$ satisfies the transformation rule
\begin{Equ}{rule-T}
  T_\beta\circ Y
  =\sum_{\bar\beta}
  Y_{\beta_1\p\bar\beta_1}\cdots Y_{\beta_M\p\bar\beta_M}
  T^*_{\bar\beta}
\end{Equ}
and $g$ the transformation rule
\begin{Equ}{rule-g}
  g_\alpha\circ Y
  =\sum_{j\geq0,\bar\alpha}
  \big(Y_{\alpha_1\p\bar\alpha_1}\cdots Y_{\alpha_N\p\bar\alpha_N}\big)_{\p j}
  T^*_{\bar\alpha j}
  +\sum_{\bar\alpha}
  Y_{\alpha_1\p\bar\alpha_1}\cdots Y_{\alpha_N\p\bar\alpha_N}
  g^*_{\bar\alpha}
\,.\end{Equ}
Here $Y\maps\RR^{n+1}\to\RR^{n+1}$ is any observer transformation,
that is with determinant~$1$.
Do to the special rule \EQU{rule-g} we define
the ``Coriolis coefficients'' $\coriolis^\beta_\alpha$ by the identity
(see \CIT{Alt2016}{}, 
for $N=1$ they are identical with the negative Christoffel symbols)
\begin{Equ}{coriolis}
  g_\alpha
  =\tx\force_\alpha+\sum_{\beta\in\{0\til n\}^{N+1}}\coriolis^\beta_\alpha T_\beta
\quad
  \tm{for}\alpha\in\{0\til n\}^N
\end{Equ}
satisfying
for all $(\alpha,\bar\gamma,\bar j)$ the transformation rule
\begin{Equ}{rule-C}\begin{arr}l
  \sum_{\gamma j}
  Y_{\gamma_1\p\bar\gamma_1}\cdots Y_{\gamma_N\p\bar\gamma_N}Y_{j\p\bar j}
  \,\coriolis^{\gamma j}_{\alpha_1\cdots\alpha_N}\circ Y
\\
  =\sum_{\bar\alpha}Y_{\alpha_1\p\bar\alpha_1}\cdots Y_{\alpha_N\p\bar\alpha_N}
  \coriolis^{*\bar\gamma\bar j}_{\bar\alpha}
  +\big(Y_{\alpha_1\p\bar\gamma_1}\cdots Y_{\alpha_N\p\bar\gamma_N}\big)_{\p\bar j}
\,,\end{arr}\end{Equ}
so that the 
so-called ``force'' $\tx\force=\seq{\tx\force_\alpha}{\alpha}$
satisfies the transformation rule 
\begin{Equ}{rule-f}
  \tx\force_\alpha\circ Y=\sum_{\bar\alpha}
  Y_{\alpha_1\p\bar\alpha_1}\cdots Y_{\alpha_N\p\bar\alpha_N}
  \tx\force^*_{\bar\alpha}
\,.\end{Equ}
Here, as said above, $Y$ is any 
observer transformation $Y:\RR^{n+1}\to\RR^{n+1}$.
With this the system \EQU{sys} reads
\begin{Equ}{sys-coriolis}
  \sum_{j\geq0}\de_{y_j}T_{\alpha j}
  -\sum_{\beta\in\{0\til3\}^{N+1}}\coriolis^\beta_\alpha T_\beta=\tx\force_\alpha
  \quad\tm{for} \alpha\in\{0\til n\}^N
\end{Equ}
were now $T$ and $\tx\force$ by \EQU{rule-T} and \EQU{rule-f}
are contravariant tensors, 
and the the Coriolis coefficients satisfy \EQU{rule-C}.
This is the general form of the system of $N$-moments.
In \CIT{Alt2016}{10 Higher moments} the following reduction has been performed.

\begin{stmt}{reduction}{Reduction}
If $\ewelt$ is the ``time vector'', the $(N-1)$-moments system
\[
  \sum_{j\geq0}\de_{y_j}T_{\gamma j}=g_\gamma
  \quad\tm{\DE{f\"ur}\EN{for}} \gamma\in\{0\til n\}^{N-1}
\]
is fulfilled for 
\[
  T_{\gamma j}:=\sum_{i\geq0}\ewelt_iT_{\gamma ij}
\,,\quad
  g_\gamma:=\sum_{i,j\geq0}\de_{y_j}\ewelt_i\cdot T_{\gamma ij}
  +\sum_{i\geq0}\ewelt_ig_{\gamma i}
\]
\end{stmt}
This gives also a reduction of the Coriolis coefficients.
\begin{prf}{} Define the test function of the $N$-moments system as
\[
  \zeta_\alpha=\zeta_\gamma\ewelt_i \quad\tm{for} \alpha=(\gamma,i)
\,.\]
That is, if $\seq{\zeta_\gamma}{\gamma}$ is a covariant tensor then
$\seq{\zeta_\alpha}{\alpha}$ is an allowed covariant test function
since $\ewelt$ is a covariant vector.
Then
\[\begin{arr}l
  0=
  \int_{\RR^4}\sum_\alpha\Big(
  \sum_j\de_{y_j}\zeta_\alpha\cdot T_{\alpha j}
  +\zeta_\alpha\, g_\alpha
  \Big)\d\leb4
\\
  =\int_{\RR^4}\sum_{\gamma i}\Big(
  \sum_j\de_{y_j}(\zeta_\gamma\ewelt_i)\cdot
  T_{\gamma ij}+\zeta_\gamma\ewelt_i g_{\gamma i}
  \Big)\d\leb4
\\
  =\int_{\RR^4}\sum_\gamma\Big(
  \sum_j\de_{y_j}\zeta_\gamma\sum_i\ewelt_iT_{\gamma ij}
  +\zeta_\gamma\sum_i\Big(
  \sum_j\de_{y_j}\ewelt_i\cdot T_{\gamma ij}+\ewelt_ig_{\gamma i}\Big)
  \Big)\d\leb4
\,,\end{arr}\]
which is the weak reduced equation.
\end{prf}

Due to examples 
we obtain the following form of the tensor $T$.
\begin{stmt}{T}{Special form of $T$}
The usual representation of the tensor $T$ is,
see for example \CIT{Alt2016}{10 Higher moments},
\begin{equ}{vv}
  T_\beta=\rho{\tx v}_{\beta_1}\cdots{\tx v}_{\beta_M}+\tilde\Pi_\beta
\,.\end{equ}
Here the ``4-velocity'' $\tx v$ is defined as in
\CIT{Alt2016}{5.2 Velocity}, that is,
as a contravariant vector $\tx v$ satisfying
\[
  {\tx v}_i\circ Y=\sum_{\bar i\geq0}Y_{i\p\bar i}\,{\tx v}^*_{\bar i}
  \quad\tm{for}i\geq0
\]
with the normalization that,
with $\ewelt$ being the covariant ``time vector'',
\[
  \sum_{i\geq0}\ewelt_i{\tx v}_i=1
\,.\]
And $\rho$ is defined as a ``spacetime mass density'',
which is an objective scalar satisfying $\rho\circ Y=\rho^*$.
Then the tensor $T$ satisfies \EQU{*.rule-T},
if $\tilde\Pi$ is also a con\-tra\-va\-ri\-ant tensor. 
\end{stmt}
Here nothing special about $\tilde\Pi$ is said,
see e.g.~the form in \EQU{theorem.*.structure}. 

%% file: SOURCES/03lagrange.tex
\sect{lagrange}{\DE{Lagrange Multiplikatoren}\EN{Lagrange multipliers}}
The aim is to derive an entropy inequality. Therefore
following Liu \& M\"uller, see the article \CIT{Liu1972}{}
and the book \CIT{Mueller1973}{}
or the books \CIT{Mueller1985}{} or \CIT{Mueller1998}{},
and also \CIT{Alt-Kontinuum}{III {\S}3},
we have to find multipliers
$\Lambda_\alpha$ for $\alpha\in\{0\til n\}^N$ 
which satisfy for ``all functions'' 
(that means for a larger set ${\cal P}'$ than
the set ${\cal P}$ of solutions of \EQU{general.*.sys})
\begin{Equ}{muller}
  \sum_{j\geq0}\de_j\tx\eta_j-\sigma
  =\sum_\alpha\Lambda_\alpha
  \big( \sum_{j\geq0}\de_{j}T_{\alpha j}-g_\alpha \big)
\,,\end{Equ}
where $\tx\eta$ is the 4-entropy and $\sigma$ the entropy production.
It is part of the entropy principle
that $\sum_{j\geq0}\de_j\tx\eta_j-\sigma$ is an
objective scalar,
hence in order to have the equation \EQU{muller} it is necessary
to state the following lemma.
This lemma and
the following is true for all values of $\seq{\Lambda_\alpha}{\alpha}$.

\begin{stmt}{Lambda}{Lemma}
For the sum
\[
  \sum_\alpha\Lambda_\alpha
  \big( \sum_{j\geq0}\de_{j}T_{\alpha j}-g_\alpha \big)
\]
being an objective scalar
it is sufficient that
$\seq{\Lambda_\alpha}{\alpha}$ is a covariant $N$-tensor.
\begin{rem}{Remark}
Here we make use of \EQU{general.*.coriolis},
that is the splitting of $g_\alpha$.
\end{rem}
\end{stmt}
\newcommand{\testfcn}{\zeta}
\begin{prf}{}
Let $\seq{\Lambda_\alpha}{\alpha}$ be a covariant $N$-tensor, that is
\[
    \Lambda^*_{\bar\alpha}=\sum_\alpha
  Y_{\alpha_1\p\bar\alpha_1}\cdots Y_{\alpha_m\p\bar\alpha_N}\Lambda_\alpha\circ Y
\,.\]
We use the splitting in \EQU{general.*.coriolis}.
Since $\seq{\tx\force_\alpha}{\alpha}$ is a contravariant $N$-tensor
it follows immediately that
\[
  \sum_\alpha\Lambda_\alpha\tx\force_\alpha
\]
is an objective scalar.
By \EQU{general.*.coriolis} it remains to consider
\[
  h_\alpha:=\sum_{j\geq0}\de_{j}T_{\alpha j}-\sum_\beta\coriolis^\beta_\alpha T_\beta
\,,\]
that is, we have to show that
\begin{Equ}{h}
  \Big(\sum_\alpha\Lambda_\alpha h_\alpha\Big)\circ Y
  =\sum_{\bar\alpha}\Lambda^*_{\bar\alpha}h^*_{\bar\alpha}
\,.\end{Equ}
If $\testfcn$ is an objective scalar, that is $\testfcn\circ Y=\testfcn^*$
hence $\de_{\bar j}\testfcn^*=\sum_jY_{j\p\bar j}\,(\de_j\testfcn)\circ Y$,
with compact support then
\[\begin{arr}l
  -\int \testfcn\sum_\alpha\Lambda_\alpha h_\alpha \d\leb4
  =\int\sum_\alpha\Big( \sum_j\de_j(\testfcn\Lambda_\alpha)T_{\alpha j}
  +\testfcn\sum_\beta\Lambda_\alpha\coriolis^\beta_\alpha T_\beta \Big)\d\leb4
\\
  =\int\Big( \sum_j\de_j\testfcn\cdot\sum_\alpha\Lambda_\alpha T_{\alpha j}
  +\testfcn\Big(\sum_{\alpha j}\de_j\Lambda_\alpha\cdot T_{\alpha j}
  +\sum_{\alpha\beta}\Lambda_\alpha\coriolis^\beta_\alpha T_\beta \Big)\Big)\d\leb4
\,.\end{arr}\]
First let us treat the last term
\[
  \sum_{\alpha j}\de_j\Lambda_\alpha\cdot T_{\alpha j}
  +\sum_{\alpha\beta}\Lambda_\alpha\coriolis^\beta_\alpha T_\beta
\,.\]
Since $\seq{\Lambda_\alpha}{\alpha}$ is a covariant $N$-tensor,
we compute for the derivatives
\[\begin{arr}l
  \de_{\bar j}\Lambda^*_{\bar\alpha}
  =\sum_{\alpha}\de_{\bar j}(Y_{\alpha_1\p\bar\alpha_1}\cdots Y_{\alpha_N\p\bar\alpha_N})
  \Lambda_\alpha\circ Y
\\\hfill
  +\sum_{\alpha j}Y_{\alpha_1\p\bar\alpha_1}\cdots Y_{\alpha_N\p\bar\alpha_N}Y_{j\p\bar j}
  \de_j\Lambda_\alpha\circ Y
\,.\end{arr}\]
Now, using \EQU{general.*.rule-C} for the Coriolis coefficients,
\[\begin{arr}l
  \sum_{\bar\alpha\bar\gamma\bar j}
  \Lambda^*_{\bar\alpha}\coriolis^{*\bar\gamma\bar j}_{\bar\alpha} T^*_{\bar\gamma\bar j}
  =\sum_{\alpha\bar\alpha\bar\gamma\bar j}\Lambda_{\alpha}\circ Y
  \,Y_{\alpha_1\p\bar\alpha_1}\cdots Y_{\alpha_N\p\bar\alpha_N}
  \coriolis^{*\bar\gamma\bar j}_{\bar\alpha} T^*_{\bar\gamma\bar j}
\\
  =\sum_{\alpha\bar\gamma\bar j\gamma j}\Lambda_{\alpha}\circ Y
  Y_{\gamma_1\p\bar\gamma_1}\cdots Y_{\gamma_N\p\bar\gamma_N}Y_{j\p\bar j}
  \,\coriolis^{\gamma j}_{\alpha_1\cdots\alpha_N}\circ Y\,T^*_{\bar\gamma\bar j}
\\\hfill
  -\sum_{\alpha\bar\gamma\bar j}\Lambda_{\alpha}\circ Y
  \de_{\bar j}(Y_{\alpha_1\p\bar\gamma_1}\cdots Y_{\alpha_N\p\bar\gamma_N})
  T^*_{\bar\gamma\bar j}
\,,\end{arr}\]
and therefore, using that $T$ is an contravariant $(N+1)$-tensor,
\[\begin{arr}l
  \sum_{\bar\alpha\bar j}\de_{\bar j}\Lambda^*_{\bar\alpha}\cdot T^*_{\bar\alpha\bar j}
  +\sum_{\bar\alpha\bar\gamma\bar j}
  \Lambda^*_{\bar\alpha}\coriolis^{*\bar\gamma\bar j}_{\bar\alpha} T^*_{\bar\gamma\bar j}
\\
  ={\sum_{\bar\alpha\bar j\alpha j}
  Y_{\alpha_1\p\bar\alpha_1}\cdots Y_{\alpha_N\p\bar\alpha_N}Y_{j\p\bar j}
  \de_j\Lambda_\alpha\circ Y\,T^*_{\bar\alpha\bar j}}
\\\qquad
  +\sum_{\alpha\bar\alpha\bar j}\Lambda_\alpha\circ Y
  \de_{\bar j}(Y_{\alpha_1\p\bar\alpha_1}\cdots Y_{\alpha_N\p\bar\alpha_N})
  T^*_{\bar\alpha\bar j}
  +\sum_{\bar\alpha\bar\gamma\bar j}
  \Lambda^*_{\bar\alpha}\coriolis^{*\bar\gamma\bar j}_{\bar\alpha} T^*_{\bar\gamma\bar j}
\\
  =\big(\sum_{\alpha j}\de_j\Lambda_{\alpha}\cdot T_{\alpha j}\big)\circ Y
\\\hfill
  +\sum_{\alpha\bar\gamma\bar j\gamma j}\Lambda_{\alpha}\circ Y
  Y_{\gamma_1\p\bar\gamma_1}\cdots Y_{\gamma_N\p\bar\gamma_N}Y_{j\p\bar j}
  \,\coriolis^{\gamma j}_{\alpha_1\cdots\alpha_N}\circ Y\,T^*_{\bar\gamma\bar j}
\\
  =\big(\sum_{\alpha j}\de_j\Lambda_{\alpha}\cdot T_{\alpha j}
  +\sum_{\alpha\gamma j}
  \Lambda_{\alpha}\coriolis^{\gamma j}_{\alpha} T_{\gamma j}
  \big)\circ Y
\,,\end{arr}\]
The term with the derivative of the test function is obviously
\[\begin{arr}l
  \sum_{\bar j}\de_{\bar j}\testfcn^*
  \sum_{\bar\alpha}\Lambda^*_{\bar\alpha} T^*_{\bar\alpha\bar j}
  =\sum_{j\bar j}Y_{j\p\bar j}\de_j\testfcn\circ Y
  \sum_{\bar\alpha}\Lambda^*_{\bar\alpha} T^*_{\bar\alpha\bar j}
\\
  =\sum_{j}\de_j\testfcn\circ Y
  \sum_{\bar\alpha}Y_{j\p\bar j}\Lambda^*_{\bar\alpha} T^*_{\bar\alpha\bar j}
  =\Big(\sum_j\de_j\testfcn\sum_\alpha\Lambda_\alpha T_{\alpha j}\Big)\circ Y
\,,\end{arr}\]
so that altogether
\[
  \int \testfcn\sum_\alpha\Lambda_\alpha h_\alpha \d\leb4
  =\int \testfcn^*\sum_{\bar\alpha}\Lambda^*_{\bar\alpha} h^*_{\bar\alpha} \d\leb4
\]
hence \EQU{h} is satisfied. 
\end{prf}


We now use the elements of the dual basis 
\[\begin{arr}l
  \{e_0'(y),e_1'(y)\til e_n'(y)\}\subset\RR^{n+1}
  \,,\tm{it is $\ewelt=e_0'$,}
\\
  \{e_0(y),e_1(y)\til e_n(y)\}\subset\RR^{n+1}
  \tm{with $e_k'\dd e_l=\kronecker{kl}$.}
\end{arr}\]
It is known that 
$\{e_1(y)\til e_n(y)\}=\welt(y)=\up{\{e'_0(y)\}}\perp$,
see \CIT{Alt2016}{3 Time and space}.
General physical statements about fluids depend only on the
vector $\ewelt(y)=e'_0(y)$ or $\welt(y)$ 
and not on single vectors $e_i(y)$, $i\geq1$ (as for example
the space directions of crystals or the director of liquid crystals).
But we are allowed to use these vectors in proofs.
In doing so we introduce values $\seq{\lambda_\gamma}{\gamma}$:
\begin{stmt}{lambda}{Definition} We define
\[\begin{arr}l
  \lambda_\gamma
  =\sum_\alpha\Lambda_\alpha e_{\gamma_1\alpha_1}\cdots e_{\gamma_N\alpha_N}
\quad\tm{or}\quad
  \Lambda_\alpha
  =\sum_\gamma\lambda_\gamma e'_{\gamma_1\alpha_1}\cdots e'_{\gamma_N\alpha_N}
\,.\end{arr}\]
The new values $\lambda_\gamma$ are  
objective scalars.
\end{stmt}
\begin{prf}{}
By this definition $\seq{\lambda_\gamma}{\gamma}$
and $\seq{\Lambda_\alpha}{\alpha}$ are equivalent quantities. 
If 
$\Lambda_\alpha$ are given as stated we conclude
\[\begin{arr}l
  \sum_\alpha\Lambda_\alpha e_{\delta_1\alpha_1}\cdots e_{\delta_N\alpha_N}
  =\sum_{\alpha,\gamma}\lambda_\gamma
  e'_{\gamma_1\alpha_1}e_{\delta_1\alpha_1}\cdots e'_{\gamma_N\alpha_N}e_{\delta_N\alpha_N} 
\\
  =\sum_{\gamma}\lambda_\gamma
  e'_{\gamma_1}\dd e_{\delta_1}\cdots e'_{\gamma_N}\dd e_{\delta_N}
  =\sum_{\gamma}\lambda_\gamma
  \kronecker{\gamma_1,\delta_1}\cdots \kronecker{\gamma_N,\delta_N}
  =\lambda_\delta
\,.\end{arr}\]
Similar the other way around. 
\end{prf}

Since we are in the proof of the main theorem
we introduce an equivalent system to the given one
presented by the terms $\sum_j\de_{y_j}T_{\alpha j}-g_\alpha$.
The new system is given by the terms 
$\sum_j\de_{y_j}T'_{\gamma j}-\rate'_\gamma$.

\begin{stmt}{reduce}{Equivalent system}
For any vectors $\seq{\Lambda_\alpha}{\alpha}$
or $\seq{\lambda_\gamma}{\gamma}$ as in \REF{lambda.} 
\[\begin{arr}c
  \sum_\alpha\Lambda_\alpha
  \big( \sum_{j\geq0}\de_{j}T_{\alpha j}-g_\alpha \big)
=
  \sum_\gamma\lambda_\gamma
  \big( \sum_{j\geq0}\de_{j}T'_{\gamma j}-\rate'_\gamma \big)
\,,\\[3.4ex]
  T'_{\gamma j}:=\sum_\alpha e'_{\gamma_1\alpha_1}\cdots e'_{\gamma_N\alpha_N}T_{\alpha j}
\quad\tm{and}\quad
  \rate'_\gamma:=\sum_\alpha e'_{\gamma_1\alpha_1}\cdots e'_{\gamma_N\alpha_N}\tx\force_\alpha
\,.\end{arr}\]
For each $\gamma$ the vector $\seq{T'_{\gamma j}}{j\geq0}$ is a covariant vector
and $\rate'_\gamma$ is an objective scalar.
\end{stmt}

\DE{Die Coriolis Kr\"afte werden durch die Definition der rechten Seite
$\rate'_\gamma$ der Differentialgleichung automatisch entfernt,
weshalb hier die Identit\"at \EQU{general.*.coriolis} so wichtig ist.
Die Coriolis Kr\"afte erscheinen nicht im Entropieprinzip.}
\EN{Since only $\tx\force_\alpha$ enter in the definition of $\rate'_\gamma$,
it means that during the process of computation
in the Liu \& M\"uller sum the fictitious forces drop out,
that is, they do not enter the entropy principle.}

\begin{prf}{}
The definition \REF{lambda.} and the
definition of $\rate'_\gamma$ implies
\[\begin{arr}l
  \sum_\alpha\Lambda_\alpha\tx\force_\alpha
 =\sum_{\alpha,\gamma}\lambda_\gamma e'_{\gamma_1\alpha_1}\cdots e'_{\gamma_N\alpha_N}
  \tx\force_\alpha
  =\sum_{\gamma}\lambda_\gamma\rate'_\gamma
\,.\end{arr}\]
And it is
\[\begin{arr}l
  \rate'_\gamma\circ Y
  =\sum_{\alpha} e'_{\gamma_1\alpha_1}\circ Y\cdots e'_{\gamma_N\alpha_N}\circ Y
  \,\force_\alpha\circ Y
\\
  =\sum_{\alpha,\bar\alpha} e'_{\gamma_1\alpha_1}\circ YY_{\alpha_1\p\bar\alpha_1}\cdots
  e'_{\gamma_N\alpha_N}\circ YY_{\alpha_N\p\bar\alpha_N}
  \,\force^*_{\bar\alpha}
\\
  =\sum_{\bar\alpha}e'^*_{\gamma_1\bar\alpha_1}\cdots e'^*_{\gamma_N\bar\alpha_N}
  \force^*_{\bar\alpha}
  =\rate'^*_\gamma
\,.\end{arr}\]
Therefore, by \EQU{general.*.coriolis}, with
\[
  h_\alpha:=\sum_j\de_{j}T_{\alpha j}-\sum_{\beta}\coriolis^\beta_\alpha T_\beta
\quad\text{and}\quad
  h'_\gamma:=\sum_j\de_{j}T'_{\gamma j}
\]
we have to show that
\begin{Equ}{h}
  \sum_\alpha\Lambda_\alpha h_\alpha=\sum_\gamma\lambda_\gamma h'_\gamma
\,.\end{Equ}
Now
\[\begin{arr}l
  \sum_{\alpha,j}\Lambda_\alpha\de_{j}T_{\alpha j}
  =\sum_{\alpha,\gamma,j}\lambda_\gamma
  e'_{\gamma_1\alpha_1}\cdots e'_{\gamma_N\alpha_N}\de_{j}T_{\alpha j}
\\
  =\sum_{\gamma,j}\lambda_\gamma\de_{j}T'_{\gamma j}
  -\sum_{\alpha,\gamma,j}\lambda_\gamma
  \de_{j}(e'_{\gamma_1\alpha_1}\cdots e'_{\gamma_N\alpha_N})T_{\alpha j}
\end{arr}\]
and
\[
  \sum_{\alpha,\beta}\Lambda_\alpha\coriolis^\beta_\alpha T_\beta
  =\sum_{\alpha,\delta,j}\Lambda_\delta\coriolis^{\alpha j}_\delta T_{\alpha j}
  =\sum_{\alpha,\gamma,j}\lambda_\gamma\Big(
  \sum_\delta e'_{\gamma_1\delta_1}\cdots e'_{\gamma_N\delta_N}
  \coriolis^{\alpha j}_\delta
  \Big)T_{\alpha j}
,\]
therefore
\[\begin{arr}l
  \sum_{\alpha}\Lambda_\alpha\Big( 
  \sum_j\de_{j}T_{\alpha j}-\sum_{\beta}\coriolis^\beta_\alpha T_\beta
  \Big)
  =\sum_{\alpha,j}\Lambda_\alpha\de_{j}T_{\alpha j}
  -\sum_{\alpha,\beta}\Lambda_\alpha\coriolis^\beta_\alpha T_\beta
\\
  =\sum_{\gamma,j}\lambda_\gamma\de_{j}T'_{\gamma j}
  -\sum_{\alpha,\gamma,j}\lambda_\gamma\Big(
  \de_{j}(e'_{\gamma_1\alpha_1}\cdots e'_{\gamma_N\alpha_N})
  +\sum_\delta e'_{\gamma_1\delta_1}\cdots e'_{\gamma_N\delta_N}
  \coriolis^{\alpha j}_\delta
  \Big)T_{\alpha j}
\end{arr}\]
and for all $(\alpha,\gamma,j)$
\begin{Equ}{term}
  \de_{j}(e'_{\gamma_1\alpha_1}\cdots e'_{\gamma_N\alpha_N})
  +\sum_\delta e'_{\gamma_1\delta_1}\cdots e'_{\gamma_N\delta_N}
  \coriolis^{\alpha j}_\delta
  =0
\,,\end{Equ}
since by the following theorem \REF{C.}
\[\begin{arr}l
  -\sum_\delta e'_{\gamma_1\delta_1}\cdots e'_{\gamma_N\delta_N}
  \coriolis^{\alpha j}_\delta
\\=
  \sum_{\delta,\beta} e'_{\gamma_1\delta_1}e_{\beta_1\delta_1}
  \cdots e'_{\gamma_N\delta_N}e_{\beta_N\delta_N}
  \de_j(e'_{\beta_1\alpha_1}\cdots e'_{\beta_N\alpha_N})
\\=
  \sum_{\beta}\kronecker{\gamma,\beta}
  \de_j(e'_{\beta_1\alpha_1}\cdots e'_{\beta_N\alpha_N})
=
  \de_j(e'_{\gamma_1\alpha_1}\cdots e'_{\gamma_N\alpha_N})
\,,\end{arr}\]
\end{prf}

\begin{stmt}{C}{Theorem} For every $(\alpha,\gamma,j)$
\[
  \coriolis^{\gamma j}_\alpha=-\sum_\beta
  e_{\beta_1\alpha_1}\cdots e_{\beta_N\alpha_N}
  \de_j(e'_{\beta_1\gamma_1}\cdots e'_{\beta_N\gamma_N})
\,,\]
since this 
is true for at least one observer.
\newline\begin{rem}{Remark} Usually true for ``inertial systems''.
\end{rem}
\end{stmt}
\begin{prf}{} The transformation rule 
for $B_\alpha^{\gamma j}:=-\coriolis_\alpha^{\gamma j}$ is
according to \EQU{general.*.rule-C}
\begin{Equ}{rule-B}\begin{arr}l
  \sum_{\bar\alpha}Y_{\alpha_1\p\bar\alpha_1}\cdots Y_{\alpha_N\p\bar\alpha_N}
  B^{*\bar\gamma\bar j}_{\bar\alpha}
  =\sum_{\gamma,j}
  Y_{\gamma_1\p\bar\gamma_1}\cdots Y_{\gamma_N\p\bar\gamma_N}Y_{j\p\bar j}
  \,B^{\gamma j}_{\alpha_1\cdots\alpha_N}\circ Y
\\\hfill
  +\big(Y_{\alpha_1\p\bar\gamma_1}\cdots Y_{\alpha_N\p\bar\gamma_N}\big)_{\p\bar j}
\,.\end{arr}\end{Equ}
Now set
\[
  B^{\gamma j}_\alpha:=\sum_\beta
  e_{\beta_1\alpha_1}\cdots e_{\beta_N\alpha_N}
  \de_j(e'_{\beta_1\gamma_1}\cdots e'_{\beta_N\gamma_N})
\,.\]
Since, see \CIT{Alt2016}{4 Change of observer}
and \REF{id.} below, 
\begin{equ}{e}
  e_{kl}\circ Y=\sum_{\bar l\geq0}Y_{l\p\bar l}e^*_{k\bar l}
  \quad\tm{for $k,l\geq0$,}
\end{equ}
\begin{equ}{eprime}
  e'^*_{k\bar l}=\sum_{l\geq0}Y_{l\p\bar l}e'^*_{kl}\circ Y
  \quad\tm{for $k,\bar l\geq0$,}
\end{equ}
\begin{equ}{kronecker}
  \sum_{m\geq0}e_{mk}e'_{ml}=\kronecker{k,l}
  \quad\tm{for $k,l\geq0$,}
\end{equ}
we compute for $(\alpha,\bar\gamma,\bar j)$
\[\begin{arr}l
  \sum_{\bar\alpha}Y_{\alpha_1\p\bar\alpha_1}\cdots Y_{\alpha_N\p\bar\alpha_N}
  B^{*\bar\gamma\bar j}_{\bar\alpha}
\\
  =\sum_{\bar\alpha,\beta}
  Y_{\alpha_1\p\bar\alpha_1}e^*_{\beta_1\bar\alpha_1}
  \cdots Y_{\alpha_N\p\bar\alpha_N}e^*_{\beta_N\bar\alpha_N}
  \de_{\bar j}(e'^*_{\beta_1\bar\gamma_1}\cdots e'^*_{\beta_N\bar\gamma_N})
\\
  =\sum_{\beta}e_{\beta_1\alpha_1}\circ Y\cdots e_{\beta_N\alpha_N}\circ Y
  \de_{\bar j}(e'^*_{\beta_1\bar\gamma_1}\cdots e'^*_{\beta_N\bar\gamma_N})
\hfill
  \qquad\text{(see \EQU{e})}
\\
  =\sum_{\beta,\gamma}e_{\beta_1\alpha_1}\circ Y\cdots e_{\beta_N\alpha_N}\circ Y
\cdot\\[-.2ex]\hfill\phantom{\sum_{\beta}}\cdot\,
  \de_{\bar j}\big(
    Y_{\gamma_1\p\bar\gamma_1}e'_{\beta_1\gamma_1}\circ Y
    \cdots Y_{\gamma_N\p\bar\gamma_N}e'_{\beta_N\gamma_N}\circ Y
  \big)
  \qquad\text{(see \EQU{eprime})}
\\
  =\sum_{\beta,\gamma}e_{\beta_1\alpha_1}\circ Y\cdots e_{\beta_N\alpha_N}\circ Y
  \,Y_{\gamma_1\p\bar\gamma_1}\cdots Y_{\gamma_N\p\bar\gamma_N}
  \de_{\bar j}(e'_{\beta_1\gamma_1}\circ Y\cdots e'_{\beta_N\gamma_N}\circ Y)
\\\hfill
  +\sum_{\beta,\gamma}
  e_{\beta_1\alpha_1}\circ Y\,e'_{\beta_1\gamma_1}\circ Y
  \cdots e_{\beta_N\alpha_N}\circ Y\,e'_{\beta_N\gamma_N}\circ Y
  \,\de_{\bar j}(Y_{\gamma_1\p\bar\gamma_1}\cdots Y_{\gamma_N\p\bar\gamma_N})
\end{arr}\]
\[\begin{arr}l
  =\sum_{\beta,\gamma}Y_{\gamma_1\p\bar\gamma_1}\cdots Y_{\gamma_N\p\bar\gamma_N}
  \,e_{\beta_1\alpha_1}\circ Y\cdots e_{\beta_N\alpha_N}\circ Y
  \de_{\bar j}(e'_{\beta_1\gamma_1}\circ Y\cdots e'_{\beta_N\gamma_N}\circ Y)
\\[-.2ex]\hfill
  +\sum_{\gamma}\kronecker{\alpha,\gamma}
  \,\de_{\bar j}(Y_{\gamma_1\p\bar\gamma_1}\cdots Y_{\gamma_N\p\bar\gamma_N})
  \qquad\text{(see \EQU{kronecker})}
\\
  =\sum_{\gamma,j}
  Y_{\gamma_1\p\bar\gamma_1}\cdots Y_{\gamma_N\p\bar\gamma_N}Y_{j\p\bar j}
  \,B^{\gamma j}_{\alpha_1\cdots\alpha_N}\circ Y
  +\de_{\bar j}\big(Y_{\alpha_1\p\bar\gamma_1}\cdots Y_{\alpha_N\p\bar\gamma_N}\big)
\,,\end{arr}\]
since
\[
  \de_{\bar j}(e'_{\beta_1\gamma_1}\circ Y\cdots e'_{\beta_N\gamma_N}\circ Y)
  =\sum_jY_{j\p\bar j}\de_j(e'_{\beta_1\gamma_1}\cdots e'_{\beta_N\gamma_N})\circ Y
\,,\]
hence $B$ satisfies \EQU{rule-B}. Therefore
the difference  
\[
  \tilde B^{\gamma j}_\alpha:=\coriolis^{\gamma j}_\alpha+\sum_\beta
  e_{\beta_1\alpha_1}\cdots e_{\beta_N\alpha_N}
  \de_j(e'_{\beta_1\gamma_1}\cdots e'_{\beta_N\gamma_N})
\]
satisfies the transformation rule
\[
  \sum_{\bar\alpha}Y_{\alpha_1\p\bar\alpha_1}\cdots Y_{\alpha_N\p\bar\alpha_N}
  \tilde B^{*\bar\gamma\bar j}_{\bar\alpha}
  =\sum_{\gamma,j}
  Y_{\gamma_1\p\bar\gamma_1}\cdots Y_{\gamma_N\p\bar\gamma_N}Y_{j\p\bar j}
  \,\tilde B^{\gamma j}_{\alpha_1\cdots\alpha_N}\circ Y
\]
which is homogeneous and therefore we can choose $\tilde B=0$.
\end{prf}

\begin{stmt}{id}{Lemma} Because
$\set{e_k}{k\geq0}$ and $\set{e'_k}{k\geq0}$ are dual basis
we know that
$\kronecker{k,l}=e'_k\dd e_l=\sum_me'_{km}e_{lm}$.
It also implies that
$\sum_me'_{mk}e_{ml}=\kronecker{k,l}$.
\end{stmt}
\begin{prf}{} Define
$E_{mk}:=e_{mk}=e_m\dd\ee_k$ and $E'_{lm}=e'_{lm}=e'_l\dd\ee_m$.
Then
\[
  \kronecker{k,l}=e'_k\dd e_l=\sum_me'_{km}e_{lm}=(E'\up{E}T)_{kl}
\,,\]
hence $E'\up{E}T=\Id$ and thus $E'(\up{E}TE'-\Id)=(E'\up{E}T)E'-E'=E'-E'=0$.
Therefore, since $E'$ is bijective, $\up{E}TE'-\Id=0$,
that is $\up{E}TE'=\Id$, which means
\[
  \kronecker{l,k}=(\up{E}TE')_{lk}=\sum_me_{ml}e'_{mk}
\,,\]
which is the assertion.
\end{prf}

%% file: SOURCES/04theorem.tex
\newcommand{\rest}{E}
\sect{theorem}{\DE{Das Entropietheorem}%
\EN{The entropy theorem}}

We start with the general system \EQU{general.*.sys} in the special case $N=2$
\begin{Equ}{sys}\begin{arr}c
  \sum_{j\geq0}\de_{y_j}T_{klj}=g_{kl}
  \quad\tm{\DE{f\"ur}\EN{for}} k,l\geq0
,\quad
  g_{kl}:=\tx\force_{kl}+\sum_\beta\coriolis^\beta_{kl} T_\beta
\end{arr}\end{Equ}
by writing the multiindex $\alpha=(k,l)$ for $k,l\geq0$,
and where all quantities are symmetric in $k$ and $l$.
The system \EQU{sys} has by definition covariant test functions,
and this is satisfied
if $T$, $\force$, and $\coriolis$ satisfy
the transformation rules which we have stated
in \EQU{general.*.rule-T},
\EQU{general.*.rule-f}, and \EQU{general.*.rule-C}. 
%
We shall consider a simple fluid which
is defined by the following representation
of the tensor components $T_{klj}$ for $k,l,j\geq0$ 
\begin{Equ}{structure}\begin{arr}l
  T_{klj}=\rho\tx v_k\tx v_l\tx v_j + E_{kl}\tx v_j + \tilde{\rm Q}_{klj}
\,,\end{arr}\end{Equ}
see \REF{general.T.}, where also the properties of the mass density $\rho$
and the 4-velocity $\tx v$ are stated.
The terms in \EQU{structure} are independent fron each other
by assuming that with the ``time vector'' $\ewelt$
\begin{Equ}{0}
  \sum_{k\geq0}\ewelt_k E_{kl}=0
\,,\quad
  \sum_{j\geq0}\ewelt_j\tilde{\rm Q}_{klj}=0
\,.\end{Equ}
The usage of the time vector $\ewelt$ says
that the ``time component'' of $E$ is zero
and that $\tilde{\rm Q}$ has no ``time derivative''.
The system \EQU{sys} therefore
can be considered as the mass-momentum-energymatrix system.

\medskip
In \REF{general.reduction.} we have defined
a reduced system of \EQU{sys} via the covariant vector $\ewelt$.
This reduced system is the mass-momentum system
\begin{Equ}{sys-mm}\begin{arr}c
  \sum_{j\geq0}\de_{y_j}T_{kj}=g_{k}
  \quad\tm{\DE{f\"ur}\EN{for}} k\geq0
\,,\\
  T_{kj}:=\sum_{l\geq0}\ewelt_lT_{klj}
  =\rho\tx v_k\tx v_j + \tilde\Pi_{kj}
\,,\quad
  \tilde\Pi_{kj}:=\sum_{l\geq0}\ewelt_l\tilde{\rm Q}_{klj}
\,,\\
  g_{k}:=\sum_{l,j\geq0}\de_{y_j}\ewelt_l\cdot T_{klj}+\sum_{l\geq0}\ewelt_lg_{kl}
\,.\end{arr}\end{Equ}
Similarly, defined as a reduction of \EQU{sys-mm} there is the mass equation
\begin{Equ}{sys-m}\begin{arr}c
  \sum_{j\geq0}\de_{y_j}T_{j}=g
\,,\\
  T_{j}:=\sum_{k\geq0}\ewelt_kT_{kj}=\rho\tx v_j + \tx\JJ_j
\,,\quad
  \tx\JJ_j:=\sum_{k\geq0}\ewelt_k\tilde\Pi_{kj}
\,,\\
  g:=\sum_{k,j\geq0}\de_{y_j}\ewelt_k\cdot T_{kj}+\sum_{k\geq0}\ewelt_kg_{k}
\,.\end{arr}\end{Equ}
Realize that we can also write
\[
  T_{j}=\sum_{k,l\geq0}\ewelt_k\ewelt_lT_{klj}
\,,\quad
  g=\sum_{k,l,j\geq0}\de_{y_j}(\ewelt_k\ewelt_l)\cdot T_{klj}
  +\sum_{k,l\geq0}\ewelt_k\ewelt_lg_{kl}
\,,\]
and that assumption \EQU{0} for $\tilde{\rm Q}$ implies
that $\tx\JJ$ and $\tilde\Pi$ satisfy
\[
  \sum_{j\geq0}\ewelt_j\JJ_{j}=0
\,,\quad  
  \sum_{j\geq0}\ewelt_j\tilde\Pi_{kj}=0 \tm{for all} k\geq0
.\]

What is left from \EQU{sys}, after one has determined the
reduced mass-momentum system \EQU{sys-mm}, is an equation
\begin{Equ}{rest}
  \sum_{j\geq0}\de_jT_{klj}^{\rest}=g^{\rest}_{kl}
  \quad\text{for $k,l\geq0$}
\,,\end{Equ}
which is given in the next statement where the vector $e_0$ satisfies 
\begin{Equ}{00}
  \GG e'_0=-\frac1{\cc^2}e_0 \,,\quad \ewelt=e'_0 \,,
\end{Equ}
see \CIT{Alt2016}{Theorem 3.4}. 
\begin{stmt}{rest}{Remaining system}
If we define the in $k$ and $l$ symmetric terms by
\[\begin{arr}l
  T^{\rest}_{klj}:=T_{klj}-e_{0k}T_{lj}-e_{0l}T_{kj}+e_{0k}e_{0l}T_j
\,,\\
  g^{\rest}_{kl}:=g_{kl}
  -\sum_{j\geq0}\de_j(e_{0k}T_{lj}+e_{0l}T_{kj})+\sum_{j\geq0}\de_j(e_{0k}e_{0l}T_j)
\,.\end{arr}\]
then system \EQU{*.rest} is fulfilled. 
For these system the reduction is zero.
\newline\begin{rem}{Remark}
There are also different representations for $g^{\rest}_{kl}$, see the proof.
\end{rem}
\end{stmt}
\begin{prf}{}
We have
\[\begin{arr}r
  \sum_{j\geq0}\de_jT^{\rest}_{klj}=\sum_{j\geq0}\de_jT_{klj}
  -\sum_{j\geq0}\de_j(e_{0k}T_{lj}+e_{0l}T_{kj})+\sum_{j\geq0}\de_j(e_{0k}e_{0l}T_j)
\\
  =g_{kl}
  -\sum_{j\geq0}\de_j(e_{0k}T_{lj}+e_{0l}T_{kj})+\sum_{j\geq0}\de_j(e_{0k}e_{0l}T_j)
  =g^{\rest}_{kl}
\,,\end{arr}\]
so that \EQU{*.rest} is satisfied.
Now, since $\ewelt=e'_0$ and $e'_0\dd e_0=1$ it follows
\[
  \sum_{k\geq0}e'_{0k}T^{\rest}_{klj}
  =\Big(\sum_{k\geq0}e'_{0k}T_{klj}-T_{lj}\Big)
  -e_{0l}\Big(\sum_{k\geq0}e'_{0k}T_{kj}-T_j\Big)
  =0
\,,\]
because by the above reduction
\[
  \sum_{k\geq0}e'_{0k}T_{klj}-T_{lj}=0
  \,,\quad
  \sum_{k\geq0}e'_{0k}T_{kj}-T_j=0
\,.\]
If we now show that for any $k$
\begin{Equ}{R}
  \sum_{l,j\geq0}\de_je'_{0l}\cdot T^R_{klj}+\sum_{l\geq0}e'_{0l}g^R_{kl}
\end{Equ}
is equal to $0$, it follows that the reduction of \EQU{*.rest} vanishes.
To prove this we write the above identity for $g^{\rest}$ as
\[\begin{arr}r
  g^{\rest}_{kl}=g_{kl}-e_{0l}g_k-e_{0k}g_l+e_{0k}e_{0l}g
\qquad\qquad\qquad\qquad\qquad\\
  -\sum_{j\geq0}\de_je_{0l}\cdot T_{kj}-\sum_{j\geq0}\de_je_{0k}\cdot T_{lj}
  +\sum_{j\geq0}\de_j(e_{0k}e_{0l})\,T_j
\,.\end{arr}\]
Using this and the above identity for $T^{\rest}_{klj}$,
making use of $e'_0\dd e_0=1$, we obtain for the term in \EQU{R}
\[\begin{arr}l
  \sum_{l,j\geq0}\de_je'_{0l}\cdot T^R_{klj}+\sum_{l\geq0}e'_{0l}g^R_{kl}
  =\sum_{l,j\geq0}\de_je'_{0l}\cdot T_{klj}+\sum_{l\geq0}e'_{0l}(g_{kl}-e_{0l}g_k)
\\\hfill
  -\sum_{l,j\geq0}e_{0k}\de_je'_{0l}T_{lj}-\sum_{l\geq0}e_{0k}e'_{0l}(g_l-e_{0l}g)
\\\hfill
  -\sum_{l,j\geq0}(\de_je'_{0l}\cdot e_{0l}T_{kj}+\de_je_{0l}\cdot e'_{0l}T_{kj})
\\\hfill
  +\sum_{l,j\geq0}
  \big(\de_je'_{0l}\cdot e_{0k}e_{0l}T_j+e'_{0l}\de_j(e_{0k}e_{0l})T_j\big)
  -\sum_{l,j\geq0}e'_{0l}\de_je_{0k}\cdot T_{lj}
\\
  =\Big(\sum_{l,j\geq0}\de_je'_{0l}\cdot T_{klj}+\sum_{l\geq0}e'_{0l}g_{kl}-g_k\Big)
  -e_{0k}\Big(\sum_{l,j\geq0}\de_je'_{0l}T_{lj}+\sum_{l\geq0}e'_{0l}g_l-g\Big)
\\\hfill
  -\de_j\Big(\sum_{l\geq0}e'_{0l}e_{0l}\Big)\cdot T_{kj}
  +\sum_{l,j\geq0}\de_j(e'_{0l}e_{0k}e_{0l})\cdot T_j
  -\sum_{l,j\geq0}\de_je_{0k}\cdot e'_{0l}T_{lj}
\\
  =\Big(\sum_{l,j\geq0}\de_je'_{0l}\cdot T_{klj}+\sum_{l\geq0}e'_{0l}g_{kl}-g_k\Big)
  -e_{0k}\Big(\sum_{l,j\geq0}\de_je'_{0l}T_{lj}+\sum_{l\geq0}e'_{0l}g_l-g\Big)
\\\hfill
  +\sum_{j\geq0}\de_je_{0k}\Big(T_j-\sum_{l\geq0}e'_{0l}T_{lj}\Big)
  =0
\,.\end{arr}\]
Hence the reduction of \EQU{*.rest} vanishes.
\end{prf}

This is a general lemma, that is, it holds without assumption \EQU{structure}.
With this assumption we perform in the next sections
\REF{exploit..} and \REF{relativ..} the entropy principle
to system \EQU{sys} and the outcome
will be that 
the physical system we derive finally will consist of
\begin{itemize}
\item the reduced mass-momentum system \EQU{sys-mm},
\item a trace of the remaining system, which will be the operation
  $\up{{\rm P}}T\up\GG{-1}{\rm P}$.
\end{itemize}
Here the map ${\rm P}$ is defined in the following lemma  
and it is important that it depends only on $\GG$ and $\ewelt$.

\begin{stmt}{P}{Lemma} We define a linear projection
${\rm P}\maps\RR^4\to\welt:=\up{\{\ewelt\}}\perp$ by 
\begin{equ}{def}
  {\rm P}=\Id \tm{on $\welt$,}\quad  {\rm P}(\GG\ewelt)=0
\,.\end{equ}
By this definition ${\rm P}$ depends only on $\GG$ and $\ewelt$.
It follows 
\[
  {\rm P}=\sum_{i\geq1}e_i\otimes e'_i
\,,\quad\text{also}\quad
  {\rm P}'=\sum_{i\geq1}e'_i\otimes e_i
\]
if we define ${\rm P}':=\up{{\rm P}}T$.
Moreover,
\begin{enum}
\num{G-1} the matrix $\up{{\rm P}}T\up\GG{-1}{\rm P}$ is
\[
  \up{{\rm P}}T\up\GG{-1}{\rm P}=\sum_{i\geq1}e'_i\otimes e'_i
\,.\]
\num{G} the matrix ${\rm P}\GG\up{{\rm P}}T$ is
\[
  {\rm P}\GG\up{{\rm P}}T=\sum_{i\geq1}e_i\otimes e_i
\,.\]
\end{enum}
\begin{rem}{Remark}
In \CIT{Alt2016}{Sec.5} we have defined $\GG^\raum={\rm P}\GG\up{{\rm P}}T$.
\end{rem}
\end{stmt}
\begin{prf}{}
Since $\welt=\fcn{span}\set{e_i}{i\geq1}$ we have
by definition ${\rm P}e_i=e_i$ for $i\geq1$.
And ${\rm P}e_0=0$ since $\GG e'_0$ and $e_0$ are proportional by \EQU{*.00}.
Since $\set{e'_k}{k\geq0}$ is the dual basis we conclude
\[
  {\rm P}=\sum_{i\geq1}e_i\otimes e'_i
\,.\]
Since $\up\GG{-1}e_i=e'_i$ for $i\geq1$, see \CIT{Alt2016}{Theorem 3.4},
we obtain
\[
  \up{{\rm P}}T\up\GG{-1}{\rm P}=\Big(\sum_{i\geq1}e'_i\otimes e_i\Big)
  \up\GG{-1}\Big(\sum_{i\geq1}e_i\otimes e'_i\Big)
  =\sum_{i\geq1}e'_i\otimes e'_i
\,,\]
and
\[
  {\rm P}\GG\up{{\rm P}}T=\Big(\sum_{i\geq1}e_i\otimes e'_i\Big)
  \GG\Big(\sum_{i\geq1}e'_i\otimes e_i\Big)
  =\sum_{i\geq1}e_i\otimes e_i
\]
since the same reads $\GG e'_i=e_i$ for $i\geq1$.
\end{prf}

\begin{stmt}{P-trans}{Transformation formula of ${\rm P}$}
It holds 
\[
  {\rm P}\circ Y\,\D{Y}=\D{Y}{\rm P}^*
\,.\]
The matrix $\up{{\rm P}}T\up\GG{-1}{\rm P}$ is covariant,
and ${\rm P}\GG\up{{\rm P}}T$ is contravariant.
\end{stmt}
\begin{prf}{}
Consider the linear map $\up{(\D{Y})}{-1}{\rm P}\circ Y\,\D{Y}$.
If a point $z^*\in\welt^*$ then the point $z\circ Y:=\D{Y}z^*$ satisfies
\[
  (z\dd\ewelt)\circ Y=(\D{Y}z^*)\dd(\ewelt\circ Y)
  =z^*\dd(\up{\D{Y}}T\ewelt\circ Y)=z^*\dd\ewelt^*=0
\]
that is $z\in\welt$. Hence ${\rm P}z=z$ and therefore
\[
  \up{(\D{Y})}{-1}{\rm P}\circ Y\,\D{Y}z^*
  =\up{(\D{Y})}{-1}({\rm P}z)\circ Y
  =\up{(\D{Y})}{-1}z\circ Y=z^*
\,.\]
Moreover, since $e_0\circ Y=\D{Y}e_0^*$, it follows from ${\rm P}e_0=0$ 
\[
  \up{(\D{Y})}{-1}{\rm P}\circ Y\,\D{Y}e_0^*
  =\up{(\D{Y})}{-1}({\rm P}e_0)\circ Y=0
\,.\]
Since the linear map is determined by these two properties
it follows $\up{(\D{Y})}{-1}{\rm P}\circ Y\,\D{Y}={\rm P}^*$.
The matrix $\up{{\rm P}}T\up\GG{-1}{\rm P}$ is covariant since
\[\begin{arr}l
  \up{{\rm P}^*}T\up{\GG^*}{-1}{\rm P}^*
  =\up{{\rm P}^*}T\up{\D{Y}}T\up{\GG}{-1}\circ Y\D{Y}{\rm P}^*
\\
  =\up{({\rm P}\circ Y\D{Y})}T\up{\GG}{-1}\circ Y{\rm P}\circ Y\D{Y}
  =\up{\D{Y}}T(\up{{\rm P}}T\up{\GG}{-1}{\rm P})\circ Y\,\D{Y}
\end{arr}\]
and the matrix ${\rm P}\GG\up{{\rm P}}T$ is contravariant since
\[\begin{arr}l
  ({\rm P}\GG\up{{\rm P}}T)\circ Y
  ={\rm P}\circ Y\D{Y}\GG^*\up{\D{Y}}T\up{({\rm P}\circ Y)}T
\\
  ={\rm P}\circ Y\D{Y}\GG^*\up{({\rm P}\circ Y\D{Y})}T
  =\D{Y}{\rm P}^*\GG^*\up{{\rm P}^*}T\up{\D{Y}}T
\end{arr}\]
for every observer transformation $Y$.
\end{prf}

For the reduced mass-momentum equation we obtain
\begin{stmt}{reduced}{Theorem}
If for \EQU{*.sys} with \EQU{*.structure}, \EQU{*.0}
the entropy principle is valid then
\begin{enum}

\num{m} the reduced mass equation becomes
\[\begin{arr}c
  \sum_{j\geq0}\de_{y_j}T_j=g
\,,\quad
  T_j:=\rho\tx v_j + \tx\JJ_j
\,.\end{arr}\]

\num{mm} the reduced mass-momentum system becomes for $k\geq0$
\[\begin{arr}c
  \sum_{j\geq0}\de_{y_j}T_{kj}=g_{k}
\,,\quad
  T_{kj}:=\rho\tx v_k\tx v_j + \tx v_k\tx\JJ_j + \tx\Pi_{kj}
\,,\\
  \tx\Pi_{kj} = p\,({\rm P}\GG\up{{\rm P}}T)_{kj} - \tx S_{kj}
\end{arr}\]
The fluxes $\tx\JJ$, $\tx\Pi$ and $\tx S$ have the property
\[\begin{arr}c
  \sum_{k\geq0}\ewelt_k\tx\Pi_{kj}=0
\,,\quad
  \sum_{k\geq0}\ewelt_k\tx S_{kj}=0
\,,\\
  \sum_{j\geq0}\ewelt_j\tx\JJ_j=0
\,,\quad
  \sum_{j\geq0}\ewelt_j\tx\Pi_{kj}=0
\,,\quad
  \sum_{j\geq0}\ewelt_j\tx S_{kj}=0
\,.\end{arr}\]
The right-hand sides $g$ and $g_k$ are as in \EQU{*.sys-m} and \EQU{*.sys-mm},
we do not say more here about these terms.
The mass equation is, of course, contained in the mass-momentum system.
\end{enum}
\end{stmt}
\begin{prf}{}
See section \REF{relativ..},
here only this:
The reduction \EQU{*.sys-mm} implies that
\[
  T_{kj}=\rho\tx v_k\tx v_j + \tilde\Pi_{kj}
\,,\quad
  \tilde\Pi_{kj}:=\sum_{l\geq0}e'_{0l}\tilde{\rm Q}_{klj}
\,,\]
a definition which is also made in
section~\REF{relativ..}, see \EQU{relativ.*.sys-mm}.
And the reduction \EQU{*.sys-m} implies that
\[
  T_{j}=\rho\tx v_j + \tx\JJ_{j}
\,,\quad
  \tx\JJ_{j}:=\sum_{k\geq0}e'_{0k}\tilde\Pi_{kj}
  =\sum_{k,l\geq0}e'_{0k}e'_{0l}\tilde{\rm Q}_{klj}
\,.\]
Now if one defines $\tx\Pi_{kj}:=\tilde\Pi_{kj}-\tx v_k\tx\JJ_j$
to have the correct formula in \REF{mm},
see the formula \EQU{relativ.*.Pi-tilde}.
And one derives
\[
  \sum_{k\geq0}e'_{0k}\tx\Pi_{kj}
  =\sum_{k\geq0}e'_{0k}(\tilde\Pi_{kj}-\tx v_k\tx\JJ_j)
  =\tx\JJ_j-\sum_{k\geq0}e'_{0k}\tx v_k\tx\JJ_j=0
\,.\]
This proves the assertion, since also
\[\begin{arr}l
  \sum_{k\geq0}e'_{0k}(({\rm P}\GG\up{{\rm P}}T))_{kj}
  =\sum_{\bar k,\bar l\geq0}\GG_{\bar k\bar l}
  \sum_{k\geq0}e'_{0k}{\rm P}_{k\bar k}{\rm P}_{j\bar l}
  =0
\,,\\
  \sum_{j\geq0}e'_{0j}(({\rm P}\GG\up{{\rm P}}T))_{kj}
  =\sum_{\bar k,\bar l\geq0}\GG_{\bar k\bar l}
  \sum_{j\geq0}e'_{0j}{\rm P}_{k\bar k}{\rm P}_{j\bar l}
  =0
\end{arr}\]
by the form of ${\rm P}$ in \REF{P.}.
That $\tx\JJ$ and $\tx\Pi$ have no ``time derivative'', that is,
\[
  \sum_{j\geq0}\ewelt_j\tx\JJ_j=0
\,,\quad
  \sum_{j\geq0}\ewelt_j\tx\Pi_{kj}=0
\,,\]
follows from \EQU{*.0} for $\tilde{\rm Q}_{klj}$.
\end{prf}

Now we perform a trace of the remaining system,
namely we multiply by the matrix $\up{{\rm P}}T\up\GG{-1}{\rm P}$.
This gives, since ${\rm P}e_0=0$ and $e'_i\dd e_0=0$ for $i\geq1$,
\[\begin{arr}l
  (\up{{\rm P}}T\up\GG{-1}{\rm P})\ddd\seq{T^{\rest}_{klj}}{k,l\geq0}
  =\sum_{i\geq1}(e'_i\otimes e'_i)\ddd\seq{T^{\rest}_{klj}}{k,l\geq0}
\\
  =\sum_{i\geq1}(e'_i\otimes e'_i)\ddd\seq{T_{klj}}{k,l\geq0}
  =\sum_{i\geq1}\sum_{k,l\geq0}e'_{ik}e'_{il}T_{klj}
\,.\end{arr}\]
Therefore the multiplication of the remaining tensor $T^\rest$
with $\up{{\rm P}}T\up\GG{-1}{\rm P}$
is the same as multiplying the original tensor $T$ with
the same matrix. We obtain

\begin{stmt}{H}{Theorem}
Multiplying the system \EQU{*.sys} by the matrix
$H:=\frac12\up{{\rm P}}T\up\GG{-1}{\rm P}$ leads to the differential equation
\[
  \sum_{j\geq0}\de_j\big(H\ddd\seq{T_{klj}}{kl}\big)=g
\,,\quad
  g:=\sum_{k,l\geq0}\Big(\sum_{j\geq0}\de_jH_{kl}\cdot T_{klj}+H_{kl}g_{kl}\Big)
\,.\]
If the assumption \EQU{*.structure} holds,
the ``total energy 4-flux'' is 
\[
  H\ddd\seq{T_{klj}}{kl}
  =\Big(\frac\rho2\,
  \tx v\dd(\up{{\rm P}}T\up\GG{-1}{\rm P})\tx v+\eps\Big)\,\tx v_j
  + \tilde q_j
  \quad\tm{for} j\geq0
,\]
where in analogy to section \REF{relativ..} the ``internal energy'' $\eps$ is
\[
  \eps:=\frac12(\up{{\rm P}}T\up\GG{-1}{\rm P})\ddd E=\frac12\up\GG{-1}\ddd E
\,.\]
and the 4-flux $\tilde q$ 
\[
  \tilde q_j=H\ddd\seq{\tilde{\rm Q}_{klj}}{kl}
  =\frac12\sum_{i\geq1}\sum_{k,l\geq0}e'_{ik}e'_{il}\tilde{\rm Q}_{klj}
\]
with $\sum_{j\geq0}\ewelt_j\tilde q_j=0$, hence 
$\tilde q$ has no time derivative.
\end{stmt}
\begin{prf}{} 
For a scalar test function $\zeta$ let 
$\zeta_{kl}:=\zeta H_{kl}$ consist of the test function
for the system \EQU{*.sys}.
It follows from \REF{P-trans.} that $H$ is a covariant tensor,
hence the test function is allowed.
Then
\[\begin{arr}l
  0=\sum_{k,l\geq0}\int_{\RR^4}\Big(
  \sum_{j\geq0}\de_j\zeta_{kl}\cdot T_{klj}+\zeta_{kl}g_{kl}\Big)
\\
  =\sum_{k,l\geq0}\int_{\RR^4}\Big(
  \sum_{j\geq0}\de_j(\zeta H_{kl})\cdot T_{klj}+\zeta H_{kl}g_{kl}\Big)
\\
  =\int_{\RR^4}\Big(\sum_{j\geq0}\de_j\zeta\cdot
  \Ubrack{\sum_{k,l\geq0}H_{kl}T_{klj}}{=H\ddd\seq{T_{klj}}{kl}}
  +\zeta\big(
  \Ubrack{\sum_{j\geq0}\de_jH_{kl}\cdot T_{klj}+\sum_{k,l\geq0}H_{kl}g_{kl}}
  {=:g}\big)
\,,\end{arr}\]
hence the new differential equation is
\[
  \sum_{j\geq0}\de_j(H\ddd\seq{T_{klj}}{kl})=g
\,,\quad
  g=\sum_{k,l\geq0}\Big(\sum_{j\geq0}\de_jH_{kl}\cdot T_{klj}+H_{kl}g_{kl}\Big)
\,,\]
where here we do not take care about $g$ in detail. 
Instead we focus here on the 4-field $\seq{H\ddd\seq{T_{klj}}{kl}}{j\geq0}$.
It is under the assumption \EQU{*.structure}
\[
  H\ddd\seq{T_{klj}}{kl}
  =\Big(\rho H\ddd(\tx v\otimes\tx v)+ H\ddd E\Big)\tx v_j
  +\sum_{k,l\geq0}H_{kl}\tilde{\rm Q}_{klj}
\,,\]
where, since $e'_0\dd\tx v=1$ and $\up\GG{-1}e_0=-\cc^2e'_0$,
\[\begin{arr}l
  2H\ddd(\tx v\otimes\tx v)=\up\GG{-1}\ddd({\rm P}\tx v\otimes {\rm P}\tx v)
  ={\rm P}\tx v\dd\up\GG{-1}{\rm P}\tx v
\\
  =(\tx v-e_0)\dd\up\GG{-1}(\tx v-e_0)
  =\tx v\dd\up\GG{-1}\tx v-2\tx v\dd\up\GG{-1}e_0+e_0\dd\up\GG{-1}e_0
\\
  =\tx v\dd\up\GG{-1}\tx v+2\cc^2\tx v\dd e'_0-\cc^2e_0\dd e'_0
\\
  =\tx v\dd\up\GG{-1}\tx v+\cc^2
  =\tx v\dd(\up\GG{-1}+\cc^2\ewelt\otimes\ewelt)\tx v
\,,\end{arr}\]
just to have a few representations of this term.
Therefore one calls the following term the ``kinetic energy''
\[
  \rho H\ddd(\tx v\otimes\tx v)
  =\frac\rho2 \tx v\dd(\up{{\rm P}}T\up\GG{-1}{\rm P})\tx v
  =\frac\rho2 {\rm P}\tx v\dd\up\GG{-1}{\rm P}\tx v
\]
and, since $H=\frac12\sum_{i\geq1}e'_i\otimes e'_i$ by \REF{P.G-1},
the ``internal energy''
\[\begin{arr}l
  \eps:= H\ddd E=\frac12(\up{{\rm P}}T\up\GG{-1}{\rm P})\ddd E
  =\frac12\sum_{i\geq1}\sum_{k,l\geq0}e'_{ik}e'_{il}E_{kl}
\\
  =\frac12\up\GG{-1}\ddd({\rm P}E\up{{\rm P}}T)=\frac12\up\GG{-1}\ddd E
\,,\end{arr}\]
since assumption \EQU{*.structure} implies ${\rm P}E=E$.
Finally for the 4-flux 
\[\begin{arr}l
  \tilde q_j:=\sum_{k,l\geq0}H_{kl}\tilde{\rm Q}_{klj}
\\
  =\frac12\sum_{i\geq1}\sum_{k,l\geq0}(e'_i\otimes e_i)_{kl}\tilde{\rm Q}_{klj}
  =\frac12\sum_{i\geq1}\sum_{k,l\geq0}e'_{ik}e'_{il}\tilde{\rm Q}_{klj}
\,.\end{arr}\]
For more about $\tilde q$ see the next statement.
\end{prf}

For the following lemma we need some formulas from section~\REF{exploit..}.
\begin{stmt}{heat}{Heat flux}
The entropy principle implies that
the 4-flux $\tilde q$ of the previous theorem has the following
representation 
\[
  \tilde q_j=\frac\rho2\,\tx v\dd(\up{{\rm P}}T\up\GG{-1}{\rm P})\tx v\,\JJ_j
  +\sum_{\bar k,k\geq0}
  \tx v_{\bar k}(\up{{\rm P}}T\up\GG{-1}{\rm P})_{\bar kk}\tx\Pi_{kj}
  +\tx q_j
\,,\]
where $\tx q$ is the ``heat flux'' occurring in the entropy production.
\end{stmt}
\begin{prf}{} From the last theorem
\[\begin{arr}l
  \tilde q_j:=
  \frac12\sum_{i\geq1}\sum_{k,l\geq0}e'_{ik}e'_{il}\tilde{\rm Q}_{klj}
  =\frac12\sum_{i\geq1}\tilde{\rm Q}'_{iij}
  \quad\text{(by \EQU{exploit.*.prime})}
\\
  =\frac12\sum_{i\geq1}
  \big( v'_iv'_i\tx\JJ_j+2\tx\Pi'_{ij}v'_i+\tx{\rm Q}'_{iij} \big)
  \quad\text{(by \EQU{exploit.L.o})}
\\
  =\frac12\sum_{i\geq1}|v'_i|^2\JJ_j
  +\sum_{i\geq1}\tx\Pi'_{ij}v'_i+\frac12\sum_{i\geq1}\tx{\rm Q}_{iij}
\,,\end{arr}\]
where the heat flux is
\[
  \tx q_j:=\frac12\sum_{i\geq1}\tx{\rm Q}_{iij}
\]
and, by the first equation in \EQU{exploit.*.prime},
\[
  \sum_{i\geq1}|v'_i|^2=\sum_{i\geq1}\sum_{k,l\geq0}e'_{ik}e'_{il}\tx v_k\tx v_l
  =\tx v\dd(\up{{\rm P}}T\up\GG{-1}{\rm P})\tx v
\,.\]
To handle the middle term we derive from \EQU{relativ.*.Pi} for $i\geq1$
\[
  \sum_{k\geq0}e'_{ik}\tx\Pi_{kj}
  =\sum_{k\geq0}e'_{ik}\sum_{\bar i\geq1}e_{\bar ik}\tx\Pi'_{\bar ij}
  =\sum_{\bar i\geq1}\big(\sum_{k\geq0}e'_{ik}e_{\bar ik}\big)\tx\Pi'_{\bar ij}
  =\tx\Pi'_{ij}
\]
and therefore
\[\begin{arr}l
  \sum_{i\geq1}\tx\Pi'_{ij}v'_i
  =\sum_{i\geq1}\big(\sum_{k\geq0}e'_{ik}\tx\Pi_{kj}\big)
  \big(\sum_{\bar k\geq0}e'_{i\bar k}\tx v_{\bar k}\big)
\\
  =\big(\sum_{i\geq1}e'_i\otimes e'_i\big)
  \ddd\seq{\tx\Pi_{kj}\tx v_{\bar k}}{k\bar k}
  =\sum_{\bar k,k\geq0}
  \tx v_{\bar k}(\up{{\rm P}}T\up\GG{-1}{\rm P})_{\bar kk}\tx\Pi_{kj}
\,.\end{arr}\]
\end{prf}

Altogether the main theorem is the 
\begin{stmt}{entropy}{Entropy theorem}
Consider the system \EQU{*.sys}, \EQU{*.structure}, \EQU{*.0}.
The application of the entropy principle
\[
  \sigma:=\sum_{j\geq0}\de_j\tx\eta_j\geq0
\]
leads to the ``mass-momentum-energy system''.
This system consists of the ``mass-momentum equation'' in \REF{reduced.mm},
and of the the ``energy equation'' in \REF{H.}, which with \REF{heat.} is
\[\begin{arr}l
  \sum_{j\geq0}\de_j\Big(
  \Big(\frac\rho2 {\rm P}\tx v\dd\up\GG{-1}{\rm P}\tx v+\eps\Big)\tx v_j
  + \tilde q_j
  \Big)=g
\,,\\
  \tilde q_j=\frac\rho2\,\tx v\dd(\up{{\rm P}}T\up\GG{-1}{\rm P})\tx v\,\JJ_j
  +\sum_{\bar k,k\geq0}
  \tx v_{\bar k}(\up{{\rm P}}T\up\GG{-1}{\rm P})_{\bar kk}\tx\Pi_{kj}
  +\tx q_j
\,.\end{arr}\]
Here the entropy and entropy 4-flux are
\[
  \eta:=\hat\eta(\rho,\eps)
\,,\quad
  \tx\eta:=\eta\tx v+\eta_{\p\rho}\tx\JJ+\eta_{\p\eps}\tx q
\,,\quad
  \eta=\ewelt\dd\tx\eta
\]
and the entropy production is
\begin{Equ}{sigma}\begin{arr}l
  0\leq\sigma=
  \sum_{k,j\geq0}\big(\sum_{i\geq1}e'_{ik}\de_j(e'_i\dd\tx v)\big)\tx S_{kj}
\\
  +\sum_{j\geq0}\de_j\hat\eta_{\p\rho}\cdot\tx\JJ_j
  +\sum_{j\geq0}\de_j\hat\eta_{\p\eps}\cdot\tx q_j
  +\hat\eta_{\p\rho}\rate^\rho
  +\hat\eta_{\p\eps}\cdot\rate^e 
\,.\end{arr}\end{Equ}
\end{stmt}
\begin{prf}{}
The proof of this theorem 
is contained in section~\REF{exploit..} and~\REF{relativ..}.
The splitting of the mass-momentum-energymatrix equation
into mass-momentum and energy equation is contained in
the statements \REF{reduced.} to \REF{heat.}.
\end{prf}

The pressure tensor $\tx\Pi$ is by \REF{reduced.mm} 
\[
  \tx\Pi = p\,{\rm P}\GG\up{{\rm P}}T - \tx S
\]
In the case of a gas $\tx\Pi = p\,{\rm P}\GG\up{{\rm P}}T$ and $\tx S=0$,
therefore the first term of the entropy production vanishes.
For fluids the stress tensor $\tx S$
has to be chosen so that the entropy production is non-negative.
This term in the entropy production is 
\[
  \sum_{k,j\geq0}\big(\sum_{i\geq1}e'_{ik}\de_j(e'_i\dd\tx v)\big)\tx S_{kj}
\]
and we show in section \REF{constit..}
that in the classical limit it converges to the well known expression
\[
  \sum_{k,j\geq1}\de_{x_j}v_k\cdot S_{kj}
\,,\]
if $\tx S$ is given by a symmetric matrix $S$. 

%% file: SOURCES/05exploit.tex
\sect{exploit}{\DE{Auswertung der Liu \& M\"uller Darstellung}%
\EN{Evaluation of the Liu \& M\"uller sum}} 
In this section we consider the relativistic moments
of up to second order, that is $N=2$ in \EQU{general.*.sys} and $\alpha=(k,l)$,
\begin{Equ}{sys}\begin{arr}c
  \sum_{j\geq0}\de_{y_j}T_{klj}-g_{kl}=0
  \quad\tm{\DE{f\"ur}\EN{for}} k,l\geq0
,\quad
  g_{kl}:=\tx\force_{kl}+\sum_{\beta\in\{0,1,2,3\}^3}\coriolis^\beta_{kl}T_\beta
\,,\end{arr}\end{Equ}
where we have set $n=3$ (the physical case).
We consider fluid equations, therefore
\begin{Equ}{structure}\begin{arr}l
  T_{klj}=\rho\tx v_k\tx v_l\tx v_j + E_{kl}\tx v_j + \tilde{\rm Q}_{klj}
\,,\end{arr}\end{Equ}
\begin{Equ}{0}
  \sum_{k\geq0}\ewelt_k E_{kl}=0
\,,\quad
  \sum_{j\geq0}\ewelt_j\tilde{\rm Q}_{klj}=0
\,.\end{Equ}
The first step in exploiting the entropy principle is to 
multiply the differential operators $\sum_{j\geq0}\de_{y_j}T_{\alpha j}-g_\alpha$
by certain factors, which Liu \& M\"uller
call Lagrange multipliers $\seq{\Lambda_\alpha}{\alpha}$,
see section~\REF{lagrange..}, and then sum up these expressions to get
\[\begin{arr}c
  \sum_\alpha\Lambda_\alpha
  \big( \sum_{j\geq0}\de_{j}T_{\alpha j}-g_\alpha \big)
\,.\end{arr}\]
Now we apply \REF{lagrange.reduce.}, that is,
we replace these sum by an equivalent system of differential operators
\begin{Equ}{Lgamma}
  L_\gamma:=\sum_{j\geq0}\de_{j}T'_{\gamma j}-\rate'_\gamma
\,,\end{Equ}
we obtain a new representation 
\[
  \sum_\alpha\Lambda_\alpha
  \big( \sum_{j\geq0}\de_{j}T_{\alpha j}-g_\alpha \big)
=
  \sum_\gamma\lambda_\gamma
  \big( \sum_{j\geq0}\de_{j}T'_{\gamma j}-\rate'_\gamma \big)
=
  \sum_\gamma\lambda_\gamma L_\gamma
\,,\]
where the quantities of the new relation are defined by  
\begin{Equ}{T-prime}\begin{arr}c
  T'_{\gamma j}:=\sum_\alpha e'_{\gamma_1\alpha_1}e'_{\gamma_2\alpha_2}T_{\alpha j}
\,,\quad
  \rate'_\gamma:=\sum_\alpha e'_{\gamma_1\alpha_1}e'_{\gamma_2\alpha_2}\force_\alpha
\,,\\
  \lambda_\gamma=\sum_\alpha\Lambda_\alpha
   e_{\gamma_1\alpha_1}e_{\gamma_2\alpha_2}
\quad\text{or}\quad
  \Lambda_\alpha=\sum_\gamma\lambda_\gamma
  e'_{\gamma_1\alpha_1}e'_{\gamma_2\alpha_2}
\,.\end{arr}\end{Equ}
Now the differential operators $L_\gamma$
have no Coriolis coefficients, therefore 
``fictitious forces'' do not appear in the entropy equation.
The representation of $T$ in \EQU{structure}
transforms by \EQU{T-prime} in the following identities 
for $k,l\geq1$ and $j\geq0$,
if one uses the assumptions in \EQU{0},
\[\begin{arr}l
  T'_{00j}=\rho\tx v_j+\tx\JJ_j
\,,\\
  T'_{k0j}=\rho v'_k\tx v_j+\tilde\Pi'_{kj}
\,,\\
  T'_{klj}=(\rho v'_kv'_l+E'_{kl})\tx v_j+\tilde{\rm Q}'_{klj}
\,,\end{arr}\]
where
\begin{Equ}{prime}\begin{arr}l
  v'_{k}:=\sum_{\bar k\geq0}e'_{k\bar k}\tx v_{\bar k}
  \tm{for}k\geq1
\,,\\
  E'_{kl}:=\sum_{\bar k,\bar l\geq0}e'_{k\bar k}e'_{l\bar l}E_{\bar k\bar l}
  \tm{for}k,l\geq1
\,,\\
  \tilde{\rm Q}'_{klj}
  :=\sum_{\bar k,\bar l\geq0}e'_{k\bar k}e'_{l\bar l}\tilde{\rm Q}_{\bar k\bar lj}
  \tm{for}k,l,j\geq0
\,,\\
  \tilde\Pi'_{kj}:=\tilde{\rm Q}'_{k0j}
  \tm{for}k\geq1
\,,\quad
  \tx\JJ_{j}:=\tilde{\rm Q}'_{00j}
\,.\end{arr}\end{Equ}
In a second step 
we show that the system \EQU{*.sys} is equivalent
to the system given by
$(L^\rho,\seq{L^v_k}{k\geq1},\seq{L^e_{kl}}{k,l\geq1})=0$, where
\begin{equ}{L=L}\begin{arr}l
  L_{00}=L^\rho
\,,\\
  L_{0k}=L_{k0}=L^v_k+v'_kL^\rho
\,,\\
  L_{kl}=L^e_{kl}+v'_kL^v_{l}+v'_lL^v_{k}+v'_kv'_lL^\rho
\end{arr}\end{equ}
for $k,l\geq1$.
These new operators are defined in the following theorem.

\begin{stmt}{L}{Theorem} Define for $k,l\geq1$ 
\[\begin{arr}l
  L^\rho:=\sum_{j\geq0}\de_j(\rho\tx v_j+\tx\JJ_j)-\rate^\rho
\,,\\
  L^v_k:=\rho\sum_{j\geq0}\tx v_j\de_jv'_k
  +\sum_{j\geq0}\de_j\tx\Pi'_{kj}+\sum_{j\geq0}\tx\JJ_j\de_jv'_k
  -\rate^v_k
\,,\\
  L^e_{kl}:=\sum_{j\geq0}\de_j(E'_{kl}\tx v_j)+\sum_{j\geq0}\de_j\tx{\rm Q}'_{klj}
  +\sum_{j\geq0}(\tx\Pi'_{lj}\de_jv'_k+\tx\Pi'_{kj}\de_jv'_l)
  -\rate^e_{kl}
\,.\end{arr}\]
Then the equations \EQU{*.L=L} are satisfied,
if for $k,l\geq1$
\begin{equ}{o}\begin{arr}l
  \tilde\Pi'_{kj}=v'_k\tx\JJ_j+\tx\Pi'_{kj}
\,,\quad
  \tilde{\rm Q}'_{klj}
  =v'_kv'_l\tx\JJ_j+\tx\Pi'_{lj}v'_k+\tx\Pi'_{kj}v'_l+\tx{\rm Q}'_{klj}
\,,\\
  \rate^v_k:=\rate'_{k0}-v'_k\rate^\rho
\,,\quad
  \rate^e_{kl}:=\rate'_{kl}-(v'_k\rate^v_l+v'_l\rate^v_k)-v'_kv'_l\rate^\rho
\,,\end{arr}\end{equ}
and of course $\rate^\rho:=\rate'_{00}$.
\end{stmt}
This follows by the same procedure as in the classical case.
\begin{prf}{}
That $L_{00}=L^\rho$ is evident. For the velocity part
\[\begin{arr}l
  L^v_k+v'_kL^\rho
  =\rho\sum_{j\geq0}\tx v_j\de_jv'_k+ v'_k\sum_{j\geq0}\de_j(\rho\tx v_j+\tx\JJ_j)
  +\sum_{j\geq0}\tx\JJ_j\de_jv'_k
\\\hfill
  +\sum_{j\geq0}\de_j\tx\Pi'_{kj} 
  -(\rate^v_k+v'_k\rate^\rho)
\\
  =\sum_{j\geq0}\de_j\big( \rho v'_k\tx v_j+v'_k\tx\JJ_j+\tx\Pi'_{kj} \big)
  -(\rate^v_k+v'_k\rate^\rho)
  =L_{0k}
\end{arr}\]
if for $k\geq1$
\[
  \tilde\Pi_{kj}=v'_k\tx\JJ_j+\tx\Pi'_{kj}
\,,\quad
  \rate'_{k0}=\rate^v_k+v'_k\rate^\rho
\,.\]
The energy part is
\[\begin{arr}l
  L^e_{kl}+v'_kL^v_{l}+v'_lL^v_{k}+v'_kv'_lL^\rho
\\
  =\sum_{j\geq0}\de_j(E'_{kl}\tx v_j)+\sum_{j\geq0}\de_j\tx{\rm Q}'_{klj}
  +\sum_{j\geq0}(\tx\Pi'_{lj}\de_jv'_k+\tx\Pi'_{kj}\de_jv'_l)
  -\rate^e_{kl}
\\\hfill
  +\rho\sum_{j\geq0}v'_k\tx v_j\de_jv'_l
  +\sum_{j\geq0}v'_k\de_j\tx\Pi'_{lj}+\sum_{j\geq0}v'_k\tx\JJ_j\de_jv'_l
  -v'_k\rate^v_l
\\\hfill
  +\rho\sum_{j\geq0}v'_l\tx v_j\de_jv'_k
  +\sum_{j\geq0}v'_l\de_j\tx\Pi'_{kj}+\sum_{j\geq0}v'_l\tx\JJ_j\de_jv'_k
  -v'_l\rate^v_k
\\
  +v'_kv'_l\sum_{j\geq0}\de_j(\rho\tx v_j)+v'_kv'_l\sum_{j\geq0}\tx\JJ_j
  -v'_kv'_l\rate^\rho
\\
  =\de_j(\rho v'_kv'_l\tx v_j)
  +\de_j(E'_{kl}\tx v_j+v'_kv'_l\tx\JJ_j+\tx\Pi'_{lj}v'_k+\tx\Pi'_{kj}v'_l
  +\tx{\rm Q}'_{klj})
\\\hfill
  -(\rate^e_{kl}+v'_k\rate^v_l+v'_l\rate^v_k+v'_kv'_l\rate^\rho)
  =L_{kl}
\end{arr}\]
if for $k,l\geq1$
\[\begin{arr}l
  \tilde{\rm Q}'_{klj}
  =v'_kv'_l\tx\JJ_j+\tx\Pi'_{lj}v'_k+\tx\Pi'_{kj}v'_l+\tx{\rm Q}'_{klj}
\,,\\
  \rate'_{kl}=\rate^e_{kl}+v'_k\rate^v_l+v'_l\rate^v_k+v'_kv'_l\rate^\rho
\,.\end{arr}\]
\end{prf}
Thus following the procedure of Liu \& M\"uller we have
for all functions 
\[\begin{arr}l
  \sum_\alpha\Lambda_\alpha\big(\sum_{j\geq0}\de_{y_j}T_{\alpha j}-g_\alpha\big)
  =\sum_\gamma\lambda_\gamma
  \big( \sum_{j\geq0}\de_{j}T'_{\gamma j}-\rate'_\gamma \big)
\\
  =\sum_\gamma\lambda_\gamma L_\gamma
  =\lambda_{00}L_{00}+\sum_{k\geq1}2\lambda_{k0}L_{k0}+\sum_{k,l\geq1}\lambda_{kl}L_{kl} 
\\
  =\lambda_{00}L^\rho+\sum_{k\geq1}2\lambda_{k0}(L^v_k+v'_kL^\rho)
\\\hfill
  +\sum_{k,l\geq1}\lambda_{kl}(L^e_{kl}+v'_kL^v_{l}+v'_lL^v_{k}+v'_kv'_lL^\rho)
\\
  =\lambda^\rho L^\rho+\sum_{k\geq1}\lambda^v_kL^v_{k}
  +\sum_{k,l\geq1}\lambda^e_{kl}L^e_{kl}
\end{arr}\]
where the new set of parameters is given by
\[\begin{arr}l
  \lambda^\rho:=\lambda_{00}+2\sum_{k\geq1}v'_k\lambda_{k0}
  +\sum_{k,l\geq1}v'_kv'_l\lambda_{kl}
\,,\\
  \lambda^v_k:=2\lambda_{k0}+2\sum_{l\geq1}v'_l\lambda_{kl}
\,,\\
  \lambda^e_{kl}:=\lambda_{kl}
\,.\end{arr}\]
for $k,l\geq1$.
Now we compute 
\[\begin{arr}l
  \lambda^\rho L^\rho+\sum_{k\geq1}\lambda^v_kL^v_{k}
  +\sum_{k,l\geq1}\lambda^e_{kl}L^e_{kl}
\\
  =\lambda^\rho\Big(\sum_{j\geq0}\de_j(\rho\tx v_j+\tx\JJ_j)-\rate^\rho\Big)
\\\hfill
  +\sum_{k\geq1}\lambda^v_k\Big(\rho\sum_{j\geq0}\tx v_j\de_jv'_k
  +\sum_{j\geq0}(\de_j\tx\Pi'_{kj}+\tx\JJ_j\de_jv'_k)-\rate^v_k\Big)
\\\hfill
  +\sum_{k,l\geq1}\lambda^e_{kl}\Big(
  \sum_{j\geq0}((\de_j(E'_{kl}\tx v_j)+\de_j\tx{\rm Q}'_{klj})
  +\sum_{j\geq0}(\tx\Pi'_{lj}\de_jv'_k+\tx\Pi'_{kj}\de_jv'_l)-\rate^e_{kl}\Big)
\\
  =\lambda^\rho\sum_{j\geq0}\de_j(\rho\tx v_j)
  +\sum_{k\geq1}\lambda^v_k\rho\sum_{j\geq0}\tx v_j\de_jv'_k
  +\sum_{k,l\geq1}\lambda^e_{kl}\sum_{j\geq0}\de_j(E'_{kl}\tx v_j)
\\\hfill
  +\sum_{j\geq0,k\geq1}\de_jv'_k\cdot\Big(
  \lambda^v_k\tx\JJ_j+2\sum_{l\geq1}\lambda^e_{kl}\tx\Pi'_{lj}\Big)
\\\hfill
  +\sum_{j\geq0}\lambda^\rho\de_j\tx\JJ_j
  +\sum_{j\geq0,k\geq1}\lambda^v_k\de_j\tx\Pi'_{kj}
  +\sum_{j\geq0,k,l\geq1}\lambda^e_{kl}\de_j\tx{\rm Q}'_{klj}
\\\hfill
  -\lambda^\rho\rate^\rho
  -\sum_{k\geq1}\lambda^v_k\rate^v_k-\sum_{k,l\geq1}\lambda^e_{kl}\rate^e_{kl}
\,,\end{arr}\]
where for the first line on the right-hand side we prove

\begin{stmt}{h}{Lemma} We can write for every function $h$ 
\[
  \sum_{j\geq0}\de_j(h\tx v_j)=
  \sum_{j\geq0}\tx v_j\de_jh
  +\sum_{j\geq0,k\geq0}\de_jv'_k\cdot(h e_{kj})
\]
\begin{rem}{Basic expression} 
For each $k\geq0$ we have the following equality $\tx\div\,e_k=0$.
This is true since the situation is connected to the standard one.
\end{rem}
\end{stmt}
\begin{prf}{}
It is for every function $h$ 
\begin{Equ}{v}
  \sum_{j\geq0}\de_j(h\tx v_j)=
  \sum_{j\geq0}\tx v_j\de_jh+h\sum_{j\geq0}\de_j\tx v_j
\,.\end{Equ}
Now since by \EQU{*.prime}
\[
  \tx v=\sum_{k\geq0}v'_ke_k \,,\quad v'_k=e'_k\dd\tx v \,,
\]
we get
\[\begin{arr}l
  \sum_{j\geq0}\de_j\tx v_j=\tx\div\,\tx v
  =\tx\div\big(\sum_{k\geq0}v'_ke_k\big)=\sum_{j\geq0,k\geq0}\de_j(v'_ke_{kj})
\\
  =\sum_{j\geq0,k\geq0}\de_jv'_k\cdot e_{kj}
  +\sum_{k\geq0}v'_k\sum_{j\geq0}\de_je_{kj}
  =\sum_{j\geq0,k\geq0}\de_jv'_k\cdot e_{kj}
\,,\end{arr}\]
since as we show now $\tx\div\,e_k=0$.
\end{prf}
\begin{prf}{of basic expression}
This follows since
$e_k\circ Y=\D{Y}e_k^*$, that is $e_k$ is a contravariant vector,
and therefore $(\tx\div\,e_k)\circ Y=\tx\div\,e_k^*$.
Since the situation is connected to the standard one we can choose $Y$
such that $e_k^*=\ee_k=\const$.
\end{prf}
From \REF{h.}, with $h$ equals $\rho$ and $E'_{\bar k\bar l}$,
we obtain for the first line on the right-hand side of our expression
\[\begin{arr}l
  \lambda^\rho\sum_{j\geq0}\de_j(\rho\tx v_j)
  +\sum_{k\geq1}\lambda^v_k\rho\sum_{j\geq0}\tx v_j\de_jv'_k
  +\sum_{\bar k,\bar l\geq1}\lambda^e_{\bar k\bar l}
  \sum_{j\geq0}\de_j(E'_{\bar k\bar l}\tx v_j)
\\
  =\lambda^\rho\sum_{j\geq0}\tx v_j\de_j\rho
  +\sum_{k\geq1}\lambda^v_k\rho\sum_{j\geq0}\tx v_j\de_jv'_k
  +\sum_{\bar k,\bar l\geq1}\lambda^e_{\bar k\bar l}\sum_{j\geq0}\tx v_j\de_jE'_{\bar k\bar l}
\\\hfill
  +\sum_{j\geq0,k\geq0}\de_jv'_k\cdot\Big(
  \lambda^\rho\rho+\sum_{\bar k,\bar l\geq1}\lambda^e_{\bar k\bar l}E'_{\bar k\bar l}
  \Big) e_{kj}
\,.\end{arr}\]
Now we can write the first three terms on the right-hand side
as a derivative of a function $\eta$,
which is later the entropy, if we let
\begin{Equ}{eta-prelim}
  \eta=\tilde\eta\big(\rho,\seq{v'_k}{k\geq1},\seq{E'_{kl}}{k,l\geq1}\big)
\,,\\
  \lambda^\rho:=\tilde\eta_{\p\rho}
\,,\quad
  \lambda^v_k:=\frac1\rho\tilde\eta_{\p v'_k}
\,,\quad
  \lambda^e_{kl}:=\tilde\eta_{\p E'_{kl}}
\,,\end{Equ}
since then by the chain rule
\[\begin{arr}l
  \sum_{j\geq0}\tx v_j\de_j\eta=
  \tilde\eta_{\p\rho}\sum_{j\geq0}\tx v_j\de_j\rho
  +\sum_{k\geq1}\tilde\eta_{\p v'_k}\sum_{j\geq0}\tx v_j\de_jv'_k
  +\sum_{\bar k,\bar l\geq1}\tilde\eta_{\p E'_{\bar k\bar l}}
  \sum_{j\geq0}\tx v_j\de_jE'_{\bar k\bar l}
\\
  =\lambda^\rho\sum_{j\geq0}\tx v_j\de_j\rho
  +\sum_{k\geq1}\lambda^v_k\rho\sum_{j\geq0}\tx v_j\de_jv'_k
  +\sum_{\bar k,\bar l\geq1}\lambda^e_{\bar k\bar l}\sum_{j\geq0}\tx v_j\de_jE'_{\bar k\bar l}
\,,\end{arr}\]
which are the first three terms on the right-hand side.
And it follows also from \REF{h.}, with $h$ equals $\eta$,
\[
  \sum_{j\geq0}\tx v_j\de_j\eta
  =\sum_{j\geq0}\de_j(\eta\tx v_j)
  -\sum_{j\geq0,k\geq0}\de_jv'_k\cdot(\eta e_{kj})
\,.\]
%
Altogether we infer that 
\[\begin{arr}l
  \sum_\alpha\Lambda_\alpha
  \big( \sum_{j\geq0}\de_{j}T_{\alpha j}-g_\alpha \big)
=
  \sum_\gamma\lambda_\gamma
  \big( \sum_{j\geq0}\de_{j}T'_{\gamma j}-\rate'_\gamma \big)
\\
  =\lambda^\rho L^\rho+\sum_{k\geq1}\lambda^v_kL^v_{k}
  +\sum_{k,l\geq1}\lambda^e_{kl}L^e_{kl}
\\
  =\sum_{j\geq0}\de_j(\eta\tx v_j)
\\\hfill
  +\sum_{j\geq0,k\geq1}\de_jv'_k
  \Big(\big(\lambda^\rho\rho
  +\sum_{\bar k,\bar l\geq1}\lambda^e_{\bar k\bar l}E'_{\bar k\bar l}-\eta\big)e_{kj}
  +\lambda^v_k\tx\JJ_j+2\sum_{l\geq1}\lambda^e_{kl}\tx\Pi'_{lj}\Big)
\\\hfill
  +\sum_{j\geq0}\lambda^\rho\de_j\tx\JJ_j
  +\sum_{j\geq0,k\geq1}\lambda^v_k\de_j\tx\Pi'_{kj}
  +\sum_{j\geq0,k,l\geq1}\lambda^e_{kl}\de_j\tx{\rm Q}'_{klj}
\\\hfill
  -\lambda^\rho\rate^\rho
  -\sum_{k\geq1}\lambda^v_k\rate^v_k-\sum_{k,l\geq1}\lambda^e_{kl}\rate^e_{kl}
\end{arr}\]
\[\begin{arr}l
  =\sum_{j\geq0}\de_j\Big(\eta\tx v_j
  +\lambda^\rho\tx\JJ_j
  +\sum_{k\geq1}\lambda^v_k\tx\Pi'_{kj}
  +\sum_{k,l\geq1}\lambda^e_{kl}\tx{\rm Q}'_{klj}\Big)
\\\hfill
  +\sum_{j\geq0,k\geq1}\de_jv'_k
  \Big(\big(\lambda^\rho\rho
  +\sum_{\bar k,\bar l\geq1}\lambda^e_{\bar k\bar l}E'_{\bar k\bar l}-\eta\big)e_{kj}
  +\lambda^v_k\tx\JJ_j+2\sum_{l\geq1}\lambda^e_{kl}\tx\Pi'_{lj}\Big)
\\\hfill
  -\sum_{j\geq0}\de_j\lambda^\rho\cdot\tx\JJ_j
  -\sum_{j\geq0,k\geq1}\de_j\lambda^v_k\cdot\tx\Pi'_{kj}
  -\sum_{j\geq0,k,l\geq1}\de_j\lambda^e_{kl}\cdot\tx{\rm Q}'_{klj}
\\\hfill
  -\lambda^\rho\rate^\rho
  -\sum_{k\geq1}\lambda^v_k\rate^v_k-\sum_{k,l\geq1}\lambda^e_{kl}\rate^e_{kl}
\\
  =\sum_{j\geq0}\de_j\tx\eta_j-\sigma
\,,\end{arr}\]
if for $j\geq0$
\begin{Equ}{eta-final}\begin{arr}l
  \eta=\tilde\eta\big(\rho,\seq{v'_k}{k\geq1},\seq{E'_{kl}}{k,l\geq1}\big)
\,,\\
  \tx\eta_j:=\eta\tx v_j
  +\lambda^\rho\tx\JJ_j
  +\sum_{k\geq1}\lambda^v_k\tx\Pi'_{kj}
  +\sum_{k,l\geq1}\lambda^e_{kl}\tx{\rm Q}'_{klj}
\,,\end{arr}\end{Equ}
and 
\begin{Equ}{sigma-final}\begin{arr}l
  \sigma:=
\\
  -\sum_{j\geq0,k\geq1}\de_jv'_k\Big(
  \big(\lambda^\rho\rho+\sum_{\bar k,\bar l\geq1}\lambda^e_{\bar k\bar l}E'_{\bar k\bar l}
  -\eta\big)e_{kj}
  +\lambda^v_k\tx\JJ_j+2\sum_{l\geq1}\lambda^e_{kl}\tx\Pi'_{lj}\Big)
\\
  +\sum_{j\geq0}\de_j\lambda^\rho\cdot\tx\JJ_j
  +\sum_{j\geq0,k\geq1}\de_j\lambda^v_k\cdot\tx\Pi'_{kj}
  +\sum_{j\geq0,k,l\geq1}\de_j\lambda^e_{kl}\cdot\tx{\rm Q}'_{klj}
\\
  +\lambda^\rho\rate^\rho
  +\sum_{k\geq1}\lambda^v_k\rate^v_k
  +\sum_{k,l\geq1}\lambda^e_{kl}\rate^e_{kl}
\,.\end{arr}\end{Equ}
Therefore for solutions of \EQU{sys}
\[
  \sum_{j\geq0}\de_j\tx\eta_j-\sigma
  =\sum_\alpha\Lambda_\alpha
  \big( \sum_{j\geq0}\de_{j}T_{\alpha j}-g_\alpha \big)
  =0
\,,\]
if the entropy quantities are given as in \EQU{eta-final}
and if $\sigma$ consists of the quantities in \EQU{sigma-final}.
For consequences see the next section. 

%% file: SOURCES/06relativ.tex
\sect{relativ}{\DE{Relativistische Entropie}%
\EN{Entropy as objective scalar}}


Here we deal with system \EQU{exploit.*.sys}
and the assumption \EQU{exploit.*.structure} and  \EQU{exploit.*.0}.
In this situation we have derived in the previous section,
that for solutions of \EQU{exploit.*.sys}
\begin{Equ}{sigma}
  \sum_{j\geq0}\de_j\tx\eta_j=\sigma
\end{Equ}
where the entropy 4-flux $\tx\eta$ satisfies \EQU{exploit.*.eta-final}
and the entropy production $\sigma$ satisfies \EQU{exploit.*.sigma-final}.
And the entropy principle $\sigma\geq0$ is required.
It is also a postulate of the entropy principle that
the equation \EQU{sigma} has to be a scalar differential equation,
which is satisfied if $\tx\eta$ is a contravariant vector
and $\sigma$ an objective scalar.
Now, the first term on the right-hand side of $\tx\eta$ in
\EQU{exploit.*.eta-final} is $\eta\tx v$
where $\tx v$ is a contravariant vector,
therefore, if $\eta$ is an objective scalar 
this term is a contravariant vector.
Remember that in \EQU{exploit.*.eta-final} we have made a constitutive relation
for $\eta$ depending on $\rho$, $v'_k=e'_k\dd\tx v$ and $E'_{kl}$.
These quantities are all objective scalars, but 
they depend on single basis vectors $e'_i$ for $i\geq1$.
This would be a non-isotropic behaviour
if $\eta$ depends really on one of these vectors. 
Such a dependence one would not allow for a simple fluid.
Therefore we come to the conclusion that $\eta$ depends only on
$\rho$ and the trace of $E'$, that is
\[
  \eps:=\frac12\sum_{k\geq1}E'_{kk}
  =\frac12\sum_{i\geq1}\sum_{\bar k,\bar l\geq0}e'_{i\bar k}e'_{i\bar l}E_{\bar k\bar l}
  =\frac12\Big(\sum_{i\geq1}e'_{i}\otimes e'_{i}\Big)\ddd E
\,,\]
which is 
the ``internal energy'' and
which of course is an objective scalar
as the sum of the objective scalars $E'_{kk}$.
It has been proved in \REF{theorem.P.}
that $\eps$ is depending on $E=\seq{E_{kl}}{k,l\geq0}$,
the energy matrix in definition \EQU{exploit.*.structure},
and apart from this only on $\GG$ and $\ewelt$, that is 
\begin{Equ}{eps}
  \eps
  =\frac12\Big(\sum_{i\geq1}e'_{i}\otimes e'_{i}\Big)\ddd E
  =\frac12(\up{{\rm P}}T\up\GG{-1}{\rm P})\ddd E
  =\frac12\up\GG{-1}\ddd E
\,,\end{Equ}
where the last equality holds by assumption \EQU{exploit.*.0} on $E$.
Thus, if the entropy $\eta$ depends only on $\rho$ and $\eps$,
then $\eta$ is an allowed objective scalar.
Therefore we assume
\begin{equ}{eta}
  \eta=\hat\eta(\rho,\eps)
\,.\end{equ}
Consequently we have for the function $\tilde\eta$ in \EQU{exploit.*.eta-final}
\[
  \tilde\eta\big(\rho,\seq{v'_k}{k\geq1},\seq{E'_{kl}}{k,l\geq1}\big)
  =\eta=\hat\eta\Big(\rho,\frac12\sum_{k\geq1}E'_{kk}\Big)
\,,\]
and it follows from \EQU{exploit.*.eta-prelim} that
\[\begin{arr}c
  \lambda^\rho=\tilde\eta_{\p\rho}=\hat\eta_{\p\rho}
\,,\quad
  \lambda^v_k=\frac1\rho\tilde\eta_{\p v'_k}=0
\,,\\
  \lambda^e_{kl}=\tilde\eta_{\p E'_{kl}}=\frac{\lambda^e}2\kronecker{k,l}
\,,\quad
  \lambda^e:=\hat\eta_{\p\eps}
\,.\end{arr}\]
With these identities and
\[
  \tx q_j:=\frac12\sum_{k\geq1}\tx{\rm Q}'_{kkj}
\,,\quad
  \rate^e:=\frac12\sum_{k\geq1}\rate^e_{kk}
\]
the formula \EQU{exploit.*.eta-final} for the entropy equation becomes
\begin{Equ}{eta-final}
  \tx\eta_j=\eta\tx v_j
  +\hat\eta_{\p\rho}\,\tx\JJ_j
  +\hat\eta_{\p\eps}\tx q_j 
\,,\quad
  \ewelt\dd\tx\eta=\eta
\,,\end{Equ}
where the last equation follows from the assumption
on $\tilde Q$ in \EQU{exploit.*.0}.
Besides this the entropy production \EQU{exploit.*.sigma-final} becomes
\[\begin{arr}c
  \sigma=
  -\sum_{j\geq0,k\geq1}\de_jv'_k\Big(
  \big(\hat\eta_{\p\rho}\rho+\hat\eta_{\p\eps}\eps-\eta\big)e_{kj}
  +\hat\eta_{\p\eps}\tx\Pi'_{kj}\Big)
\\\hfill
  +\sum_{j\geq0}\de_j\hat\eta_{\p\rho}\cdot\tx\JJ_j
  +\sum_{j\geq0}\de_j\hat\eta_{\p\eps}\cdot\tx q_j
\\\hfill
  +\hat\eta_{\p\rho}\rate^\rho
  +\hat\eta_{\p\eps}\cdot\rate^e 
\,.\end{arr}\]

To proceed further let us assume that, 
in analogy to the classical case,  
\begin{Equ}{theta}
  \frac1\theta:=\hat\eta_\eps(\rho,\eps)>0
\,,\quad
  \frac\mu\theta:=\hat\eta_\rho(\rho,\eps)
\,.\end{Equ}
Here $\theta$ is the ``absolute temperature'' and 
$\mu$ the ``chemical potential''.
We define the preliminary version
$\seq{\tx S'_{kj}}{j\geq0,k\geq1}$ of the stress tensor by
\[\begin{arr}l
  \tx S'_{kj}:=
  \theta\big(\big(\eta-\hat\eta_{\p\rho}\rho-\hat\eta_{\p\eps}\eps\big)e_{kj}
  -\hat\eta_{\p\eps}\tx\Pi'_{kj}\big)
  =\big(\theta\eta-\mu\rho-\eps\big)e_{kj} - \tx\Pi'_{kj}
\end{arr}\]
for $k\geq1$, so that one gets 
for the entropy production the final version
\begin{Equ}{sigma-final}\begin{arr}c
  0\leq\sigma=
  \hat\eta_{\p\eps}\sum_{j\geq0,k\geq1}\de_jv'_k\tx S'_{kj}
  +\sum_{j\geq0}\de_j\hat\eta_{\p\rho}\cdot\tx\JJ_j
  +\sum_{j\geq0}\de_j\hat\eta_{\p\eps}\cdot\tx q_j
\\\hfill
  +\hat\eta_{\p\rho}\rate^\rho
  +\hat\eta_{\p\eps}\cdot\rate^e 
\end{arr}\end{Equ}
where $\sigma\geq0$ by the entropy principle.
If we now define the ``pressure'' $p$ by
\begin{Equ}{p}
  p:=\theta\eta-\mu\rho-\eps
\,,\end{Equ}
which is Gibbs relation, 
the above definition takes the common form
\begin{Equ}{Pi-prime}
 \tx\Pi'_{kj}=p\, e_{kj} - \tx S'_{kj}
 \quad\tm{for} k\geq1
\end{Equ}
We have to write this in terms of the reduced mass-momentum system
\EQU{theorem.*.sys-mm}
\begin{Equ}{sys-mm}\begin{arr}c
  \sum_{j\geq0}\de_{y_j}T_{kj}=g_{k}
  \quad\tm{\DE{f\"ur}\EN{for}} k\geq0
\,,\\
  T_{kj}:=\sum_{l\geq0}\ewelt_lT_{klj}
  =\rho\tx v_k\tx v_j + \tilde\Pi_{kj}
\,,\quad
  \tilde\Pi_{kj}:=\sum_{l\geq0}\ewelt_l\tilde{\rm Q}_{klj}
\,.\end{arr}\end{Equ}
Now, by \EQU{exploit.*.prime}, for $k\geq0$
\[
  \tilde{\rm Q}'_{k0j}
  =\sum_{\bar k,\bar l\geq0}e'_{k\bar k}e'_{0\bar l}\tilde{\rm Q}_{\bar k\bar lj}
  =\sum_{\bar k\geq0}e'_{k\bar k}\tilde\Pi_{\bar kj}
\]
or, by renaming $k$ as $\bar k$ and vice versa,
\[
  \tilde{\rm Q}'_{\bar k0j}
  =\sum_{k\geq0}e'_{\bar kk}\tilde\Pi_{kj}
\]
hence for $k\geq0$, making use of \REF{lagrange.id.},
\[\begin{arr}l
  \tilde\Pi_{kj}=\sum_{\bar k\geq0}e_{\bar kk}\tilde{\rm Q}'_{\bar k0j}
  =e_{0k}\tx\JJ_j+\sum_{i\geq1}e_{ik}\tilde\Pi'_{ij}
  \quad\text{(using \EQU{exploit.*.prime})}
\\
  =e_{0k}\tx\JJ_j+\sum_{i\geq1}e_{ik}(v'_i\tx\JJ_j+\tx\Pi'_{ij})
  \quad\text{(using \EQU{exploit.L.o})}
\\
  =\Big(e_{0k}+\sum_{i\geq1}e_{ik}v'_i\Big)\tx\JJ_j
  +\sum_{i\geq1}e_{ik}\tx\Pi'_{ij}
  =\tx v_{k}\tx\JJ_j
  +\sum_{i\geq1}e_{ik}\tx\Pi'_{ij}
\,.\end{arr}\]
Therefore, if we define for $k\geq0$
the ``pressure tensor'' and the ``stress tensor'' by
\begin{Equ}{Pi}
  \tx\Pi_{kj}:=\sum_{i\geq1}e_{ik}\tx\Pi'_{ij}
\quad\text{and}\quad
  \tx S_{kj}:=\sum_{i\geq1}e_{ik}\tx S'_{ij}
\,,\end{Equ}
we have shown
\begin{Equ}{Pi-tilde}
  \tilde\Pi_{kj}=\tx v_{k}\tx\JJ_j+\tx\Pi_{kj}
  \tm{for} k\geq0
,\end{Equ}
and the identity \EQU{Pi-prime} becomes
\[\begin{arr}l
  \tx\Pi_{kj}:=\sum_{i\geq1}e_{ik}\tx\Pi'_{ij}
  =\sum_{i\geq1}e_{ik}\big(p\, e_{ij} - \tx S'_{ij}\big)
\\
  =p\sum_{i\geq1}e_{ik}e_{ij}-\tx S_{kj}
  =p\,({\rm P}\GG\up{{\rm P}}T)_{kj}-\tx S_{kj}
  \quad\text{(using \REF{theorem.P.G})}
\,,\end{arr}\]
that is, the well known formula
\begin{Equ}{formula}
 \tx\Pi = p\,{\rm P}\GG\up{{\rm P}}T - \tx S
\,.\end{Equ}

This shows \REF{theorem.reduced.mm}, 
and therefore the statements about the
reduced mass-mo\-men\-tum system are proved.
We come back to the entropy production $\sigma$ in \EQU{sigma-final},
which is not so final since it contains the term
\[
  \sum_{j\geq0,k\geq1}\de_jv'_k\tx S'_{kj}
  =\sum_{j\geq0}\Big(\sum_{i\geq1}\de_jv'_i\tx S'_{ij}\Big)
\]
depending on $\tx S'=\seq{\tx S'_{ij}}{i\geq1,j\geq0}$ and not on
the stress tensor $\tx S=\seq{\tx S_{kj}}{k,j\geq0}$.
Now, we get from the definition \EQU{Pi} 
\[
  \sum_{k\geq0}e'_{ik}\tx S_{kj}
  =\sum_{\bar i\geq1}\sum_{k\geq0}e'_{ik}e_{\bar ik}\tx S'_{\bar ij}
  =\sum_{\bar i\geq1}\kronecker{i,\bar i}\tx S'_{\bar ij}
  =\tx S'_{ij}
\]
and thus
\begin{Equ}{fluid}\begin{arr}r
  \sum_{j\geq0}\Big(\sum_{i\geq1}\de_jv'_i\tx S'_{ij}\Big)
=
  \sum_{k,j\geq0}\Big(\sum_{i\geq1}e'_{ik}\de_jv'_i\Big)\tx S_{kj}
\\=
  \sum_{k,j\geq0}\Big(\sum_{i\geq1}e'_{ik}\de_j(e'_i\dd\tx v)\Big)\tx S_{kj}
\,.\end{arr}\end{Equ}
That this is the generalization of the term
in the classical case is shown in the next session.

%% file: SOURCES/07constit.tex
\sect{constit}{\DE{Konstitutive Gleichung}%
\EN{Constitutive equation for fluids}}
\renewcommand{\bar}{\overline}
\newcommand{\VVV}{{\bf V}}
\newcommand{\QQQ}{{\bf Q}}
We deal with the term \EQU{relativ.*.fluid} for the stress tensor
$\tx S=\seq{\tx S_{kj}}{k,j\geq0}$
\begin{Equ}{fluid}
  \sum_{k,j\geq0}\big(\sum_{i\geq1}e'_{ik}\de_j(e'_i\dd\tx v)\big)\tx S_{kj}
\,.\end{Equ}
which is part of the entropy inequality $\sigma\geq0$ in \REF{theorem.entropy.}.
We show that this expression converges as $\cc\to\infty$ to the well known term
of the Navier-Stokes limit.
%
By this limit we mean that
$e'_k\to\bar e'_k$ and $e_k\to\bar e_k$ as $\cc\to\infty$,
where the limit basis are given as usual:

\begin{stmt}{basis}{Limit basis} We obtain in the standard case the limits 
\[
  \bar e'_0=\begin{mat}1\\0\end{mat}
\,,\quad
  \bar e_0=\begin{mat}1\\\VVV\end{mat}
\,,\quad
  \bar e_i=\begin{mat}0\\\QQQ\ee_i\end{mat}
\,,\quad
  \bar e'_i=\begin{mat}-\VVV\dd\QQQ\ee_i\\\QQQ\ee_i\end{mat}
  \quad\tm{for} i\geq1
\]
where $\D_{x}\VVV$ is antisymmetric and $\QQQ$ depends only on $t$.
\end{stmt}
\begin{prf}{} We consider the standard case, that is,
we assume that $|\bar e'_0|=1$. Then
\[
  \ewelt=e'_0\to\bar\ewelt=\bar e'_0=\begin{mat}1\\0\end{mat}
\quad\text{with}\quad
  \welt=\up{\{e'_0\}}\perp\to\up{\{\bar e'_0\}}\perp=:\bar\welt
\,,\]
which implies, since $\bar e'_0\dd\bar e_0=1$, that
\[
  e_0\to\bar e_0=\begin{mat}1\\\VVV\end{mat}=:\tx\VVV
\]
which is the definition of the vector $\VVV$.
The elements $\set{\bar e_i}{i\geq1}$ are
an or\-tho\-nor\-mal set of $\bar\welt$, that is
\[
  \bar e_i=\begin{mat}0\\\QQQ\ee_i\end{mat} \tm{for} i\geq1
\]
which is the definition of the orthonormal matrix $\QQQ$.
Then the representation of the elements $\bar e'_i$ follow easily.
Now with $Y$ being a Newton transformation,
$\QQQ$ satisfies the transformation rule
\[
  \begin{mat}0\\\QQQ\circ Y\,\ee_i\end{mat}=\bar e_i\circ Y
  =\D{Y}\bar e_i^*=\begin{mat}1\tab0\\\dot X\tab Q\end{mat}
  \begin{mat}0\\\QQQ^*\ee_i\end{mat}
  =\begin{mat}0\\Q\,\QQQ^*\ee_i\end{mat}
\]
that is $\QQQ\circ Y=Q\,\QQQ^*$. Hence if $\QQQ^*$ is the Identity for
at least one ${}^*$-observer, then $\QQQ\circ Y$ is a function of $t^*$ only
and so $\QQQ$ is independent of $x$.
Similarly, $\VVV$ satifies the transformation rule
\[
  \begin{mat}1\\\VVV\circ Y\end{mat}=\bar e_0\circ Y
  =\D{Y}\bar e_0^*=\begin{mat}1\tab0\\\dot X\tab Q\end{mat}
  \begin{mat}1\\\VVV^*\end{mat}
  =\begin{mat}1\\\dot X + Q\,\VVV^*\end{mat}
\]
that is $\VVV\circ Y=\dot X + Q\,\VVV^*$, and therefore
\[
  \sum_{j\geq1}Q_{j\bar j}(\de_{x_j}\VVV_i)\circ Y=\de_{x^*_{\bar j}}(\VVV_i\circ Y)
  =\dot Q_{i\bar j}+\sum_{\bar i\geq1}Q_{i\bar i}\de_{x^*_{\bar j}}\VVV_{\bar i}^*
\,,\]
hence
\[
  (\de_{x_j}\VVV_i)\circ Y=(\dot Q\up{Q}T)_{ij}
  +\sum_{\bar i,\bar j\geq1}Q_{i\bar i}Q_{j\bar j}\de_{x^*_{\bar j}}\VVV_{\bar i}^*
\,.\]
It follows that if $\VVV^*$ is zero for at least one ${}^*$-observer,
then $\seq{\de_{x_j}\VVV_i}{ij}$ is antisymmetric.
\end{prf}
Since
\[
  \sum_{k\geq0}e'_{0k}\tx S_{kj}=0
\,,\quad
  \sum_{j\geq0}e'_{0j}\tx S_{kj}=0
\,,\]
we have also in the classical limit
\[
  0=\sum_{k\geq0}\bar e'_{0k}\tx S_{kj}=\tx S_{0j}
\,,\quad
  0=\sum_{j\geq0}\bar e'_{0j}\tx S_{kj}=\tx S_{k0}
\,,\]
therefore
\[
  \tx S=\begin{mat}0\tab0\\0\tab S\end{mat}
\,.\]  
Having this in mind we compute,
since $\tx v=(1,v)$ and
\[
  \begin{mat}-\VVV\dd\QQQ\ee_i\\\QQQ\ee_i\end{mat}\dd\tx v
  =(v-\VVV)\dd\QQQ\ee_i=\big(\up\QQQ{T}(v-\VVV)\big)_i
\,,\]
and since $\QQQ$ depends only on $t$,
\[\begin{arr}l
  \sum_{k,j\geq0}\big(\sum_{i\geq1}e'_{ik}\de_j(e'_i\dd\tx v)\big)\tx S_{kj}
\longrightarrow
  \sum_{k,j\geq0}\big(\sum_{i\geq1}\bar e'_{ik}\de_j(\bar e'_i\dd\tx v)\big)\tx S_{kj}
\\=\Big(
  \sum_{i\geq1}\begin{mat}-\VVV\dd\QQQ\ee_i\\\QQQ\ee_i\end{mat}\otimes
  \tx\grad\Big(\begin{mat}-\VVV\dd\QQQ\ee_i\\\QQQ\ee_i\end{mat}\dd\tx v\Big)
  \Big)\ddd\tx S
\\=\Big(\sum_{i\geq1}
  \begin{mat}-\VVV\dd\QQQ\ee_i\\\QQQ\ee_i\end{mat}\otimes
  \tx\grad\big(\big(\up\QQQ{T}(v-\VVV)\big)_i\big)
  \Big)\ddd\tx S
\\=\Big(\sum_{i\geq1}
  (\QQQ\ee_i)\otimes\grad\big(\big(\up\QQQ{T}(v-\VVV)\big)_i\big)
  \Big)\ddd S
\\=
  \sum_{k,j\geq1}\sum_{i\geq1}(\QQQ\ee_i)_k
  \de_{x_j}\big(\big(\up\QQQ{T}(v-\VVV)\big)_iS_{kj}
\\=
  \sum_{k,j\geq1}\sum_{i\geq1}\QQQ_{ki}
  \de_{x_j}\big(\sum_{l\geq1}\QQQ_{li}(v-\VVV)_l\big)S_{kj}
\\=
  \sum_{k,j\geq1}\sum_{l\geq1}\big(\sum_{i\geq1}\QQQ_{ki}
  \QQQ_{li}\big)\de_{x_j}(v-\VVV)_l\cdot S_{kj}
=
  \sum_{k,j\geq1}\de_{x_j}(v-\VVV)_k\cdot S_{kj}
\\=
  \sum_{k,j\geq1}\big(\de_{x_j}v_k-\de_{x_j}\VVV_k\big) S_{kj}
=
  \sum_{k,j\geq1}\de_{x_j}v_k\cdot S_{kj}
\,,\end{arr}\]
if $S$ is symmetric.
This is true since $\seq{\de_{x_j}\VVV_k}{jk}$ is antisymmetric.

%% file: SOURCES/refer-thermo.tex
\newcommand{\authors}[1]{#1:}
\newcommand{\titles}[1]{\emph{#1}.}
\newcommand{\sources}[1]{#1.}
\newcommand{\published}[1]{ #1}
\newcommand{\electro}[1]{\\#1}

%% file: TITELEI/dynamo-AMSA.bbl
\begin{thebibliography}{99}

\bibitem{Alt-FluidSolid}
\authors{H.W. Alt}
\titles{Entropy principle and interfaces. Fluids and Solids}
\sources{Advances in Mathematical Sciences and Applications (AMSA),
Vol. 19, pp. 585-663}
\published{2009}

\bibitem{Alt-Kontinuum}
\authors{H.W. Alt}
\titles{Mathematical Continuum Mechanics}
\sources{Script lecture TUM M\"unchen 2011-2017}

\bibitem{Alt2016}
\authors{H.W. Alt}
\titles{Relativistic equations for the Chapman-Enskog hierarchy}
\published{Advances in Mathematical Sciences and Applications (AMSA),
Vol. 25, pp.131-179, 2016}

\bibitem{DeGrootMazur}
\authors{S.R. de Groot, P. Mazur} 
\titles{Non-Equilibrium Thermodynamics}
\published{North-Hol\-land 1962}

\bibitem{HutterJohnk}
\authors{Kolumban Hutter, Klaus J\"ohnk}
\titles{Continuum Methods of Physical Modeling.
Continuum Mechanics, Dimensional Analysis, Turbulence}
\published{Springer-Verlag 2004}

\bibitem{Liu1972}
\authors{I-Shih Liu}
\titles{Method of Lagrange multipliers for exploitation
of the entropy principle}
\sources{Arch. Rat. Mech. Anal. 46, pp.131-148}
\published{1972}

\bibitem{LandauLifschitz2}
\authors{L.D. Landau, E.M. Lifschitz}
\titles{Lehrbuch der theoretischen Physik. Band II. Klassische Feldtheorie}
\sources{12. Auflage}
\published{Akademie-Verlag Berlin 1992}

\bibitem{LandauLifschitz6}
\authors{L.D. Landau, E.M. Lifschitz}
\titles{Lehrbuch der theoretischen Physik. Band VI. Hydrodynamik}
\sources{3. Auflage}
\published{Akademie-Verlag Berlin 1974}

\bibitem{Mueller1973}
\authors{Ingo M\"uller}
\titles{Thermodynamik. Grundlagen der Materialtheorie}
\published{Bertelsmann Universit\"atsverlag 1973}

\bibitem{Mueller1985}
\authors{Ingo M\"uller}
\titles{Thermodynamics}
\published{Pitman 1985}

\bibitem{Mueller1998}
\authors{Ingo M\"uller, Tommaso Ruggeri}
\titles{Rational Extended Thermodynamics}
\sources{$2^{\rm nd}$ Edition, Springer Tracts in Natural Philosophy, Vol. 37}
\published{Springer 1998}

\bibitem{Prigogine1954}
\authors{I. Prigogine, R. Defay}
\titles{Chemical Thermodynamics}
\published{Longmans Green 1954}

\bibitem{Truesdell1969}
\authors{C. Truesdell}
\titles{Rational Thermodynamics}
\sources{Second Edition}
\published{Springer-Verlag New York 1984}

\bibitem{TruesdellNoll}
\authors{C. Truesdell, W.Noll} 
\titles{Non-Linear Field Theories of Mechanics}
\sources{Third Edition Springer-Verlag}
\published{2004}


\end{thebibliography}
